\documentclass[journal]{IEEEtran}

\usepackage{graphicx}
\usepackage{balance}
\usepackage{comment}
\usepackage{cite}
\usepackage{amssymb}
\usepackage[tight,footnotesize]{subfigure}
\usepackage[active]{srcltx}
\usepackage{amsmath}

\newcommand{\norm}[1]{\left\lVert#1\right\rVert}
\graphicspath{{./figs/}}
\renewcommand{\baselinestretch}{0.95}
\usepackage{eurosym}
\usepackage[plain]{algorithm}
\usepackage{algorithmic}
\usepackage{multicol}
\usepackage{dsfont}
\usepackage{amsfonts}

\usepackage{cases}
\usepackage{xcolor}
\IEEEoverridecommandlockouts

\begin{document}

\title{Distributed Search Planning in 3D Environments with a Dynamically Varying  Number of Agents}

\author{Savvas~Papaioannou,~Panayiotis~Kolios,~Theocharis~Theocharides,\\~Christos~G.~Panayiotou~ and ~Marios~M.~Polycarpou

\IEEEcompsocitemizethanks{\IEEEcompsocthanksitem The authors are with the KIOS Research and Innovation Centre of Excellence (KIOS CoE) and the Department of Electrical and Computer Engineering, University of Cyprus, Nicosia, 1678, Cyprus.\protect\\
E-mail:\texttt{\{papaioannou.savvas, pkolios, ttheocharides, christosp, mpolycar\}@ucy.ac.cy}}
\thanks{}}

\markboth{IEEE Transactions on Systems, Man, and Cybernetics: Systems, (EARLY ACCESS) doi: 10.1109/TSMC.2023.3240023}%
{Papaioannou \MakeLowercase{\textit{et al.}}: Distributed Search Planning in 3D Environments with a Dynamically Varying  Number of Cooperative Agents}

\IEEEcompsoctitleabstractindextext{%
\begin{abstract}
In this work, a novel distributed search-planning framework is proposed, where a dynamically varying team of autonomous agents cooperate in order to search multiple objects of interest in 3D. It is assumed that the agents can enter and exit the mission space at any point in time, and as a result the number of agents that actively participate in the mission varies over time. The proposed distributed search-planning framework takes into account the agent dynamical and sensing model, and the dynamically varying number of agents, and utilizes model predictive control (MPC) to generate cooperative search trajectories over a finite rolling planning horizon. This enables the agents to adapt their decisions on-line while considering the plans of their peers, maximizing their search planning performance, and reducing the duplication of work. 
\end{abstract}
}

\maketitle

\IEEEdisplaynotcompsoctitleabstractindextext
\IEEEpeerreviewmaketitle

\section{Introduction} \label{sec:Introduction}

\IEEEPARstart{I}{n} emergency response situations the immediate deployment of the response team is imperative for saving people's lives. In such situations the ability to plan and organize predictable, precise, and efficient cooperative searches of the affected area is of the highest importance in order to locate people in danger.  In general, an emergency response mission can be divided into two main tasks \cite{UNCHR2} i.e., assessment, and search-and-rescue. In the assessment task, the rescue team first assesses the damages and hazards of the affected region and then determines the areas that need to be searched for locating survivors or people in need. During the assessment task the rescue team organizes and plans the search mission. Subsequently, the purpose of the search-and-rescue task is to perform organized, complete and efficient searches of the affected area in order to locate survivors and provide rescue. 

We envision that a team of distributed autonomous mobile agents (i.e., unmanned aerial vehicles (UAVs) or drones), capable of conducting optimized and coherent search-planning in 3D, can significantly enhance the capabilities and success rate of the rescue team in emergency response situations. The assessment task is captured in this work through a mission pre-planning step in which the affected area that needs to be searched, once identified, is decomposed into a number of artificial cells according to the UAV's sensing capabilities and required search-effort. Then we propose a distributed search-planning framework in which multiple autonomous agents cooperate in order to efficiently search the affected area. 
In our previous works \cite{Savvas_2021a,Savvas_2021b} we have presented a novel planning framework for the problem of  3D search planning with a single autonomous agent. Therefore, the motivation of this article is to design a multi-agent distributed 3D search-planning framework with improved performance and more capabilities compared to the single agent case.

In this work, we propose a distributed search-planning framework, based on model predictive control (MPC) \cite{Christofides2013}, for the problem of cooperative searching in 3D environments with a dynamically varying number of agents. In particular, in this work, it is assumed that the agents can enter and exit the mission space (i.e., to recharge their depleted batteries) at any point in time, and as a result the number of active agents that participate in the mission changes over time. This necessitates the need for efficient planning and cooperation amongst the team of agents, so that they can adapt their plans and make decisions on-line in order to better accommodate the collective objective of the team. 

More specifically, the objective is for a dynamically varying team of agents to cooperate in order to efficiently search multiple objects of interest in 3D (i.e., the total surface area of each object of interest must be searched) with certain detection probability. The agents are equipped with a camera-based sensing system with finite field-of-view (FoV), which they use to scan the surface of the objects of interest while maintaining the required detection probability (specified at the beginning of the mission). Therefore, the agents cooperate in order to scan the total surface area of each object of interest, searching for survivors while trying to minimize the duplication of work. To achieve this, it is assumed that the agents can opportunistically communicate and exchanging their search-maps and their future intentions whenever they are in communication range with each other. The proposed approach does not require any form of coordination between the agents, thus enabling them to plan their decisions autonomously and in parallel with each other, while optimizing the collective objective of the team. Overall, the contributions of this work are as follows:
\begin{itemize}
    \item We propose a novel distributed search planning framework, based on model predictive control (MPC), which enables a dynamically varying number of autonomous agents to cooperatively search in 3D multiple objects of interest, without requiring any form of coordination.
    \item We derive a mixed integer quadratic programming (MIQP) mathematical formulation for the  distributed 3D search planning problem which can be solved efficiently with widely available optimization tools. 
    \item Finally, the performance of the proposed approach is demonstrated through a series of qualitative and quantitative synthetic experiments. 
\end{itemize}

The rest of the paper is organized as follows. Section~\ref{sec:Related_Work} summarizes the related work on search-planning and coverage control with multiple agents. Section \ref{sec:system_model} develops the system model and Section \ref{sec:area_decomp} discusses the mission pre-planning step which takes place prior to search-planning. Then, Section \ref{sec:multi_agent} discusses the details of the proposed distributed multi-agent 3D search planning framework, Sec. \ref{sec:Evaluation} evaluates the proposed framework and finally Sec. \ref{sec:conclusion} concludes the paper.

\section{Related Work}\label{sec:Related_Work}
In the recent years, we have witnessed an unprecedented interest in UAV-based applications and automation technologies \cite{UAV2,J3,j2,Wang2018,Savvas2021,UAV4,j1,TAES2022,ICUAS2022}, with particular interest in planning techniques. In this work the problem of trajectory planning with the objective of searching an area of interest with multiple agents is investigated.
An interesting work on this topic is shown in \cite{Berger2015}, where the authors proposed a centralized formulation for the problem of multi-agent search-planning which they solve using mixed-integer linear programming (MILP). The work in \cite{Ivan2007} proposes a two-stage centralized-assignment, decentralized-covering algorithm in which the area of interest is first divided into non-overlapping regions in a centralized fashion, and then assigned to the UAV agents. Each UAV agent then runs a local covering algorithm to search its assigned area. In a similar fashion, the work in \cite{Yao2017}, proposes a hierarchical cooperative planning framework for finding a target in a 2D environment. In \cite{Yao2017} the area of interest is first decomposed and prioritized into subregions and then allocated to the UAV agents. Each UAV agent then uses a local receding horizon controller (RHC) for searching its allocated area. In \cite{Hussein2014} a centralized market-based multi-robot task allocation algorithm is proposed for assigning regions of interest to mobile agents. The idea of distributed task allocation for multi-agent search operations is illustrated in \cite{Zhao2016}. Multi-agent search-planning is also investigated in \cite{San2018}, where the authors evaluate various discrete search-planning algorithms. The problem of distributed search-planning is investigated in \cite{Tisdale2009}, with the goal of searching and localizing a stationary ground target with a team of UAVs. The authors propose a distributed control framework for maximizing the probability of target detection with a team of UAVs over a finite planning horizon. This method however, requires coordination between the agents and works in a sequential fashion.

In \cite{Grancharova2015} a distributed trajectory planning approach is proposed based on linear model predictive control (MPC), where multiple UAVs are guided with the goal of forming a communication network around multiple targets. More recently, the authors in \cite{Niu2022} proposed a decentralized MPC approach for multi-UAV trajectory planning for obstacle avoidance, whereas in \cite{Wang2021} a consensus algorithm for distributed cooperative formation trajectory planning is proposed based on artificial potential fields and consensus theory. In \cite{Du2021} a sampling-based chance-constrained 2D trajectory planning approach is proposed for multiple UAV agents with probabilistic geo-fencing constraints, whereas in \cite{Ahmed2021} a particle-swarm optimization (PSO) approach is proposed for distributed collision-free trajectory planning with a team of UAVs operating in stochastic environments. More recently, the authors in \cite{Mou2021} have proposed a deep reinforcement learning based 3D area coverage approach with a swarm of UAV agents, whereas in \cite{Tolstaya2021} a multi-robot coverage approach is proposed based on  spatial graph neural networks. Moreover, in \cite{xu2022multi} the authors investigate the problem of full coverage search with multiple agents in cluttered environments, and finally, the work in \cite{TCYB1} proposes a distributed sweep coverage algorithm for multi-agent systems in uncertain environments.

In comparison with the related works above, in this work we propose a distributed search-planning approach which does not require the commonly used two-stage procedure of centralized-assignment and decentralized coverage. Instead, in the proposed approach the agents cooperatively decide in a rolling-horizon fashion which regions of interest to visit and how to visit them, generating search-plans online, thus tackling the overall search-planning problem in a distributed fashion. In addition, in contrast with the aforementioned  literature, in this work we consider a dynamically varying number of agents which a) exhibit limited sensing and communication capabilities, and b) are prone to random battery failures.

Other related works on this topic include the problem of adaptive and role-based collaboration in multi-agent systems which is investigated in \cite{Zhu2012,Zhu2015}. The authors propose a mathematical model (i.e., E-CARGO) which can be used to describe in a rigorous mathematical way, relationships and interactions within a typical multi-agent system; thus enabling the design and implementation of efficient algorithms for various real-world problems of multi-agent systems \cite{Zhu2015}. Moreover, the work in \cite{Ta2014} presents a factor graph optimization framework to tackle the problems of estimation and optimal control jointly, whereas the work in \cite{Mukadam2018} poses the problem of continuous-time motion-planning as a probabilistic inference problem with Gaussian processes, and then proposes efficient gradient-based optimization planners (GPMP). More recently, the problem of online motion planning is investigated in \cite{Alwala2021} with joint sampling and trajectory optimization over factor graphs. Factor-graph based motion planning techniques (e.g., \cite{Ta2014,Alwala2021}) are mostly concerned with the determination of a single obstacle-free trajectory between the starting and goal locations. In contrast, in this work we propose a rolling-horizon distributed search-planning approach which allows a dynamically varying number of autonomous agents (governed by dynamical, sensing, communication, and battery constraints) to decide their control inputs, and generate cooperative trajectories in order to search in 3D multiple objects of interest. Finally, the proposed approach is formulated as a convex MIQP which can be solved optimally using existing optimization solvers, whereas graph-based motion planning methods (e.g., \cite{Mukadam2018}) often rely on iterative gradient-based optimization methods which they do not offer any global optimality guarantees.

\color{black}

\section{System Model} \label{sec:system_model}

\subsection{Agent Dynamics} \label{ssec:agent_dynamics}
A team $\mathcal{M} = \{1,\ldots,|\mathcal{M}|\}$ of autonomous mobile agents (i.e., UAVs), is deployed inside a bounded surveillance region $\mathcal{A}$. Each agent $j \in \mathcal{M}$ evolves in 3D space according to the following discrete-time linear dynamical model:
\begin{equation} \label{eq:agent_dynamics}
    x^j_{t+1} = \Phi x^j_{t} + \Gamma u^j_{t} - \Gamma u_g, ~\forall j\in \mathcal{M}
\end{equation}
where $x^j_t = [\text{x}^j,\dot{\text{x}}^j]_t^\top \in \mathbb{R}^6$ denotes the agent's state at time $t$ which consists of position $\text{x}^j_t=[p_x, p_y, p_z]_t \in \mathcal{A} \subset \mathbb{R}^3$ and velocity $\dot{\text{x}}^j_t = [\nu_x,\nu_y,\nu_z]_t \in \mathbb{R}^3$ components in 3D cartesian coordinates. The agent can be controlled by applying an amount of force $u^j_t  \in \mathbb{R}^3$ in each dimension, thus $u^j_{t} = [\text{u}_x, \text{u}_y, \text{u}_z]_{t}^\top$ denotes the applied force vector at $t$ and the constant $u_g = [0, 0, m^j g]^\top$ denotes the force of gravity where $g = 9.81 \text{m}/\text{s}^2$ is the Earth's gravitational acceleration and $m^j$ is the agent mass. The matrices $\Phi$ and $\Gamma$ are given by:
\begin{equation}
\Phi = 
\begin{bmatrix}
    \text{I}_{3\times3} & \Delta T \cdot \text{I}_{3\times3}\\
    \text{0}_{3\times3} & \phi \cdot \text{I}_{3\times3}
   \end{bmatrix},~
\Gamma = 
\begin{bmatrix}
    \text{0}_{3\times3} \\
     \gamma \cdot \text{I}_{3\times3}
   \end{bmatrix}
\end{equation}

\noindent where $\Delta T$ is the sampling interval, $\text{I}_{3\times3}$ is the identity matrix of dimension $3 \times 3$ and $\text{0}_{3\times3}$ is the zero matrix of dimension $3 \times 3$. The parameters $\phi$ and $\gamma$ are further given by $\phi =  (1-\eta)$ and $\gamma = \frac{\Delta T}{m^j}$, and the parameter $\eta \in [0,1]$ is used to model the air resistance.

\subsection{Agent Battery and Communication model} \label{ssec:battery}

Each agent $j \in \mathcal{M}$ exhibits a nominal flight time of $\mathcal{T}^j$ time-steps which depends on the agent's onboard battery lifetime. However, the agent's onboard battery health deteriorates due to irreversible physical and chemical changes that take place with usage and aging, which makes the nominal flight time inaccurate due to imprecise battery state-of-charge cycle calculations. For this reason agent's $j$ battery can be depleted during a mission at some time $t \le \mathcal{T}^j$ with probability $p^j_b(t)$. When this happens, the agent needs to exit the mission space and return to its ground station (relying on backup power) located at $\mathcal{G}^j \in \mathbb{R}^3$ for recharging. We model the battery depletion event of agent's $j$ battery at time $t$, as a Bernoulli random variable $\mathcal{B}^j \in \{0,1\}$ with conditional probability distribution given by:
\begin{equation}\label{eq:battery_eq}
    Pr(\mathcal{B}^j = 1|t) = p^j_b(t) = \frac{1}{1+\alpha_1^j e^{-\beta_1^j(t-\alpha_1^j)}}
\end{equation}

\noindent where the parameters $\alpha_1^j$ and $\beta_1^j$ control the severity of the battery's aging. 
Due to the random battery depletion events that occur during the mission, only a subset $\tilde{\mathcal{M}}_t \subseteq \mathcal{M}$ of agents actively participate in the search-planning task at any given time instance $t$. Moreover, we assume that the recharging time $t_R$ is distributed uniformly in the interval $[\mathcal{T}^\text{start}_R,\mathcal{T}^\text{stop}_R]$, i.e., $t_R \sim \mathcal{U}(\mathcal{T}^\text{start}_R,\mathcal{T}^\text{stop}_R)$ for all agents. Thus, after $t_R$ time-steps of recharging, the agents can enter again the mission space, and continue their mission. To achieve some form of cooperation, the set of agents that participate in the mission $\tilde{\mathcal{M}}_t \subseteq \mathcal{M}$, exchange information whenever they are in communication range. An agent $j \in \tilde{\mathcal{M}}_t$ can communicate and receive information from the group of neighboring agents $\mathcal{N}_t^j = \{i \ne j \in \tilde{\mathcal{M}}_t : \norm{Hx^i_t - Hx^j_t}_2 \le C_R \}$ where $H$ is a matrix which extracts the position coordinates from the agent's state vector i.e., $H x_t =\text{x}_t=[p_x, p_y, p_z]^\top_t$ and $C_R$ is the communication range which we assume in this work to be the same for all agents.


\subsection{Agent Sensing Model} \label{ssec:sensing_model}

Each agent is equipped with a camera system which is used for acquiring snapshots of the objects of interest. Assuming that the camera field-of-view (FoV) angles in the horizontal and vertical axis are equal, the projection of the camera FoV on a planar surface is given by a square with side length $r$ as $r(d) = 2 d  \tan\left(\frac{\varphi}{2}\right)$,
where $d$ denotes the distance in meters between the location of the agent and the surface of the object that needs to be searched, and $\varphi$ is the angle opening of the FoV according to the camera specifications. Before taking a snapshot of the object of interest the agent first aligns its camera with respect to the surface in such a way so that the optical axis of the camera (i.e., the viewing direction) is parallel to the normal vector of the face. An object of interest is searched when its total surface area is included in the agents's acquired images. The acquired images are then processed by a computer vision module to detect people. The quality of the acquired images depends on the distance between the agent and the object of interest. Therefore, the probability of detecting people $p_d(d)$ in the acquired images depends on the distance $d$ between the agent and the object of interest according to:
\begin{equation} \label{eq:pd}
p_d(d) = 
\begin{cases} 
   0 & , ~\text{if }~ d \le d_\text{min} \\
   \text{max} (0, ~ 1-\frac{d-d_\text{min}}{d_\text{max}-d_\text{min}}) & ,~ \text{if }~ d > d_\text{min}
\end{cases}
\end{equation}

\noindent where $d_\text{min}$ and $d_\text{max}$ are the minimum and maximum camera working distance for detecting people in the acquired frames. Although in this work we are utilizing a simplified detection probability model to demonstrate the proposed search planning framework, more realistic  sensor detection models \cite{Popovic2020,Thrun2002} can be incorporated without requiring any changes in the problem formulation.



\subsection{Objects of Interest and Obstacles Model} \label{ssec:cuboids}
The objects of interest that need to be searched and the obstacles in the environment that need to be avoided by the agents are represented in this work by rectangular cuboids of various sizes (referred to hereafter as cuboids). A rectangular cuboid is a convex hexahedron in three dimensional space which exhibits six rectangular faces (i.e., where each pair of adjacent faces meets in a right angle).

A point $x \in \mathbb{R}^{3}$ belongs to the cuboid $\mathcal{C}$ (i.e., $x \in \mathcal{C}$) if the linear system of inequalities $A x \le B$ is satisfied, where  $A$ is a $6 \times 3$ matrix, with each row corresponding to the outward normal vector $\alpha_i = [\alpha_{i,x},\alpha_{i,y},\alpha_{i,z}]$ of the plane which contains the $i_\text{th}$ face of the cuboid and $B = [b_1,\ldots,b_6]^\top$ is a $6 \times 1$ column vector, where each element $b_i$ is obtained by the dot product between $\alpha_i$ and a known point on plane $i$. For the rest of the paper, we will use the matrices $A$ and $B$ to denote the system of linear inequalities $A x \le B$ associated with the rectangular cuboid $\mathcal{C}$.

\section{Mission Pre-planning} \label{sec:area_decomp}

The amount of search-effort which is required in order to successfully search an object of interest, during an emergency response mission, is generally determined during the mission assessment phase \cite{UNCHR2} which is conducted at the mission control, prior to the search mission. During the assessment phase, the rescue team assesses the situation at hand (e.g., potential hazards, missing people, importance of the object, etc.) and specifies the amount of search-effort required for conducting an efficient and effective search.  In this work, the search-effort is captured by the  detection probability (i.e., Eqn. \eqref{eq:pd}) issued at the central station, before the mission begins. A high detection probability allows for detailed and accurate snapshots of the object of interest. However, the size of the FoV inversely decreases with the amount of detail captured in the acquired images, and as such more snapshots are needed to cover the whole surface of the object with a high detection probability. 

\begin{figure}
	\centering
	\includegraphics[width=\columnwidth]{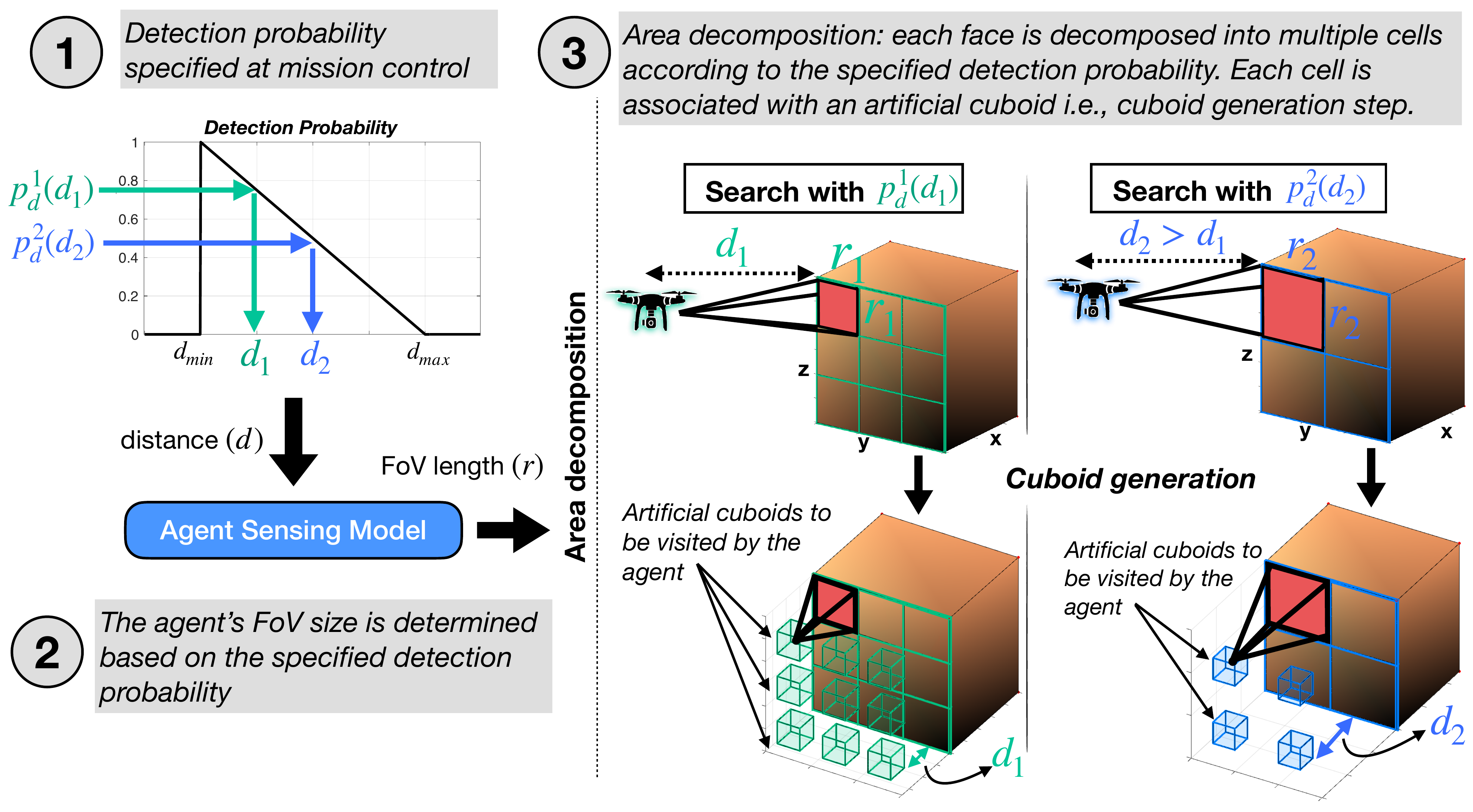}
	\caption{The figure illustrates the mission pre-planning step.}	
	\label{fig:are_decomp}
	\vspace{-5mm}
\end{figure}

 In order to allow the UAV agent to search the total surface area of the object of interest with the specified detection probability, the area around the object is decomposed into multiple cuboids as illustrated in Fig. \ref{fig:are_decomp}. In essence, once the distance $d$ between the agent and the object of interest is determined according to the specified detection probability $p_d(d)$, the agent's FoV footprint $r \times r$ is computed according to the agent's sensing model, as illustrated by step 2 in Fig. \ref{fig:are_decomp}. Subsequently, each face of the object of interest is decomposed into cells of size $r \times r$, forming a 3D grid, as illustrated by step 3 in the figure. For each cell, an artificial cuboid is generated and placed at distance $d$ from the center of the cell as depicted in Fig. \ref{fig:are_decomp}. Then, by guiding the UAV agent through all the generated cuboids, we make sure that the total surface area of each face is searched according to the specified detection probability. This is because, once the agent resides within a particular cuboid, the projected camera FoV on the face's surface captures the area of the corresponding cell as illustrated in Fig. \ref{fig:are_decomp}.
The area decomposition process discussed above, allows us to transform the 3D search problem into an optimal control problem i.e., finding the UAV control inputs, such that the agent is guided through all the generated cuboids in an optimal way. 

\section{Multi-Agent 3D Search Planning} \label{sec:multi_agent}
In this section we develop a rolling-horizon distributed model predictive control (DMPC) algorithm \cite{Negenborn2014,Christofides2013} for the cooperative guidance of a team of UAV agents with the purpose of searching in 3D multiple objects of interest while avoiding collisions with the obstacles in the environment. Our DMPC formulation, does not require any explicit coordination between the UAV agents and thus allows the agents to operate independently and in parallel with each other. In the proposed approach the agents can enter and exit the mission space according to the condition of their batteries and opportunistically cooperate with each other by exchanging information whenever they are in communication range, optimizing their future search plans. In particular, the agents in communication range exchange a) state information i.e., their current location, b) their search-maps, c) their future search plans and finally d) their flight time. Based on the exchanged information the agents seek to optimize their future plans in order to minimize the duplication of work and improve the search efficiency.

\subsection{Centralized Control} \label{ssec:centralized}
Let us assume that the mission control has issued the required detection probability and that the mission pre-planning step discussed in Sec. \ref{sec:area_decomp} is completed, i.e., the faces of the object of interest that need to be searched are covered with a total of $N$ artificial cuboids $\mathcal{C}_n, n \in [1,..,N]$.

We assume that a centralized station is in place, where all the necessary information is being transmitted, and which in turn decides the control actions for each agent by solving the 3D search planning problem jointly among all agents. The centralized formulation of the multi-agent 3D search planning problem is shown in problem (P1), where we seek to obtain the optimal joint control inputs for all agents i.e., $u^j_{t|t},\ldots,u^j_{t+T-1|t}, \forall j \in \mathcal{M}$ over a rolling planning horizon of length $T$ time-steps, by solving an open-loop optimal control problem, with the goal of guiding the agents to visit all cuboids while ensuring that the work is not duplicated i.e., a cuboid is not searched by more than one agent. Once the sequence of joint control inputs is found, the first control inputs $u^j_{t|t}, \forall j$ in the sequence are executed by the agents and the procedure described above is repeated for the subsequent time-steps.\color{black}

\begin{algorithm}
\begin{subequations}
\begin{align}
&\hspace*{-7.5mm}\textbf{Problem (P1)}: \texttt{Centralized MPC} &  \nonumber\\
& \hspace*{-7.5mm}~~~\underset{\mathbf{U_t},\mathbf{Y}}{\min} ~\mathcal{J}_\text{centralized}(\mathbf{X_t}, \mathbf{U_t}, \mathbf{Y}) &\label{eq:objective_P1} \\
&\hspace*{-7.5mm}\textbf{subject to} ~ j \in \{1,..,|\mathcal{M}|\}, \tau \in [0,\ldots,T-1] \textbf{:}  &\nonumber\\
&\hspace*{-7.5mm} x^j_{t+\tau+1|t} = \Phi x^j_{t+\tau|t} + \Gamma u^j_{t+\tau|t} - \Gamma u_g & \hspace*{-30mm} \forall \tau, j \label{eq:P1_1}\\
&\hspace*{-7.5mm} x^j_{t|t} = x^j_{t|t-1} & \hspace*{-30mm}\forall j \label{eq:P1_2}\\
&\hspace*{-7.5mm} Hx^j_{t+\tau+1|t} \notin \mathcal{C}_\psi & \hspace*{-30mm} \forall \tau, \psi, j \label{eq:P1_3}\\
&\hspace*{-7.5mm} x^j_{t+\tau+1|t} \in \mathcal{X} &\hspace*{-30mm} \forall \tau, j \label{eq:P1_4}\\
&\hspace*{-7.5mm} |u^j_{t+\tau+1|t}| \le u_\text{max} &\hspace*{-30mm} \forall \tau, j \label{eq:P1_5}\\
&\hspace*{-7.5mm} A_{n,l} H x^j_{t+\tau+1|t} + (M-B_{n,l}) b^j_{\tau,n,l} \le M & \hspace*{-30mm}\forall \tau,n,l \label{eq:P1_x1}\\
&\hspace*{-7.5mm} L \tilde{b}^j_{\tau,n} - \sum_{l=1}^L b^j_{\tau,n,l} \le 0 & \hspace*{-30mm}\forall \tau,n \label{eq:P1_x2}\\
&\hspace*{-7.5mm}  \hat{b}^j_{n} \le \sum_\tau \tilde{b}^j_{\tau,n} & \hspace*{-30mm}\forall n \label{eq:P1_x3}\\
&\hspace*{-7.5mm} y^j_{n} \le \hat{b}^j_{n} + V_t^j(n) + \sum_{i \ne j \in \mathcal{M}} \left[ V_t^i(n) + P_t^i(n) \right] &\hspace*{-30mm} \forall n, j \label{eq:P1_6}\\
&\hspace*{-7.5mm} y^j_n, b^j_{\tau,n,l}, \tilde{b}^j_{\tau,n}, \hat{b}^j_n,V_t^j(n),P_t^j(n) \in \{0,1\} &\hspace*{-30mm} \forall n, j \label{eq:P1_7}
\end{align}
\vspace{-5mm}
\end{subequations}
\end{algorithm}

\subsubsection{Objective Function}
\noindent In problem (P1) we are interested in optimizing a system-wide objective function i.e., Eqn. \eqref{eq:objective_P1} over the planning horizon of length $T$ time-steps for the joint controls over all $j \in \mathcal{M}$ agents. In (P1), the bold capital letters indicate quantities over all agents thus $\mathbf{X_t} = \{X^1_t,..,X_t^{|\mathcal{M}|}\}$ is the combined state for all $|\mathcal{M}|$ agents with $X^j_t = \{x^j_{t+\tau+1|t}\}, \forall \tau \in [0,..,T-1]$, where the notation $x_{t+\tau+1|t}$ is used here to denote the future (i.e., planned) agent state at time $t+\tau+1$ based on the current time-step $t$. Similarly, $\mathbf{U_t} = \{U^1_t,..,U_t^{|\mathcal{M}|}\}$ denotes the agent joint mobility controls inputs with $U^j_t = \{u^j_{t+\tau|t}\}, \forall \tau$ and finally $\mathbf{Y} = \{Y^1,..,Y^{|\mathcal{M}|}\}$ are binary variables indicating whether a specific artificial cuboid $\mathcal{C}_n$ has been visited or will be visited in the future by some agent $j$ with $Y^j = \{y^j_1,..,y^j_n\}, n \in [1,..,N]$.

In essence our goal is to find the agent joint control inputs $u^j_{t+\tau|t}, \forall \tau, j$ which will maximize the number of cuboids that will be visited during the planning horizon. The objective function can thus be defined as $\underset{\mathbf{U_t},\mathbf{Y}}{\min} ~\mathcal{J}_\text{centralized}(\mathbf{X_t}, \mathbf{U_t}, \mathbf{Y})=$
\begin{align} \label{eq:centralized_mission_objective}
    &\underset{\mathbf{U_t},\mathbf{Y}}{\min}~~ w_1 \sum_{j=1}^{\mathcal{M}} \|Hx^j_{t+\tau^\star+1|t}-x_j^{\star}\|^2_2 +\\ \notag &
    w_2 \sum_{j=1}^{\mathcal{M}} \sum_{\tau=1}^{T-1} \|u^j_{t+\tau|t}-u^j_{t+\tau-1|t}\|^2_2 - w_3 \sum_{j=1}^{\mathcal{M}}\sum_{n=1}^{N}y^j_{n}  &
\end{align}

\noindent where $w_i > 0$ are tuning weights, $\tau^\star \in [0,..,T-1]$, and $x^\star_j$ is the centroid of the nearest unvisited cuboid to agent's $j$ current location. This is computed as $x^\star_j = c(\mathcal{C}_{n_j^\star})$ where $n_j^\star$ is given by: $ n^\star_j = \underset{n \in \tilde{N}^j_t}{\arg\min} \|Hx^j_{t+\tau^\star+1|t} - c(\mathcal{C}_n)\|_2$, 
\noindent with $\tilde{N}^j_t$ denoting agent's $j$ set of all unvisited cuboids and the operator $c(\mathcal{C}_n)$ returns the centroid of cuboid $\mathcal{C}_n$. Therefore, the first term in Eqn. \eqref{eq:centralized_mission_objective} guides all agents towards their nearest unvisited cuboids. The second term minimizes the deviations between consecutive control inputs over all agents in order to produce smooth trajectories which the UAV agents can follow and finally, the last term maximizes the number of cuboids to be visited by the team of agents over the planning horizon, indicated by the binary variable $y^j_n$ which is defined as: $y^j_{n} = 1 \implies \exists \tau \in [0,..,T-1] : Hx^j_{t+\tau+1|t} \in \mathcal{C}_n$ 

\subsubsection{Constraints}
Eqn. \eqref{eq:P1_1} and Eqn. \eqref{eq:P1_2} are due to the agent dynamical model assuming a known initial state $x^j_{t|t}$. Then, Eqn. \eqref{eq:P1_3} defines the obstacle avoidance constraints of agent $j$ with all obstacles $\mathcal{C}_\psi, \psi \in [1,..,\Psi]$ in the environment where $\Psi$ denotes the total number of obstacles present and $\mathcal{C}_\psi, \psi \in [1,..,\Psi]$ denotes the cuboid representation of obstacle $\psi$. We can now say that agent $j$ avoids a collision with an obstacle $\psi \in \Psi$ when: $Hx^j_{t+\tau+1|t} \notin \mathcal{C}_\psi,~\forall \psi \in \Psi, \forall \tau \in \{0,\ldots,T-1\}$, which can be implemented with the following constraints:
\begin{subequations}
\begin{align}
    &A_{\psi,l} (H x^j_{t+\tau+1|t}) > B_{\psi,l} - M z^j_{\tau,\psi,l} &\hspace*{1mm} \forall \tau,\psi, l \label{eq:obstacle1}\\
    &\sum_{l=1}^L z^j_{\tau,\psi,l} \le L-1 &\hspace*{1mm} \forall \tau,\psi \label{eq:obstacle2}
\end{align}
\end{subequations}

\noindent In Eqn. \eqref{eq:obstacle1}, $A_{\psi,l}$ and $B_{\psi,l}$ define the coefficients of the equation of the plane which contains the $l_\text{th}$ face of the obstacle. When the system of linear inequalities $A_{\psi,l} (H x^j_{t+\tau+1|t}) < B_{\psi,l}, \forall l \in [1,..,L]$ is true then the agent is contained within the obstacle, which signifies that a collision has occurred. Thus a collision is avoided when $\exists l \in \{1,\ldots,L\}: A_{\psi,l} H x^j_{t+\tau+1|t} > B_{\psi,l}$. This is achieved a) with the binary variable $z^j_{\tau,\psi,l}$ which counts the number of times the inequality $A_{\psi,l} H x^j_{t+\tau+1|t} > B_{\psi,l}$ is violated for agent $j$, regarding the face $l$ of obstacle $\psi$ and b) with the constraint in Eqn. \eqref{eq:obstacle2} which makes sure that the number of  violations is less than $L-1$ where $L=6$ is the total number of faces that compose the obstacle. In Eqn. \eqref{eq:obstacle1}, $M$ denotes a large positive constant, also known as big-$M$ \cite{Bosch2005}, which is selected in such a way so that the constraint shown in Eqn. \eqref{eq:obstacle1} is satisfied at all times when $z^j_{\tau,\psi,l}=1$.

The Eqn. \eqref{eq:P1_4} constrains the agent's state within the bounded set $\mathcal{X}$, and the constraint in Eqn. \eqref{eq:P1_5} limits the values of the control input within the range $[-u_\text{max},+u_\text{max}]$ as shown. The constraints in Eqn. \eqref{eq:P1_x1}-\eqref{eq:P1_x3} determine whether agent $j$ resides inside cuboid $\mathcal{C}_n$ at time $\tau$ (relative to the horizon) via the binary variables $b^j_{\tau,n,l}, \tilde{b}^j_{\tau,n}$ and $\hat{b}^j_{n}$. Thus, the constraints in Eqn. \eqref{eq:P1_x1}-\eqref{eq:P1_x3} allow the agent to search in 3D an object of interest by passing through all artificial cuboids that have been generated for this object. The $n_\text{th}$ cuboid $\mathcal{C}_n$ is visited by the agent when the system of linear inequalities $A_{n,l} H x^j_{t+\tau+1|t} < B_{n,l}, \forall l$ holds for every face $l$. Thus the binary variable $b^j_{\tau,n,l}$ indicates whether this inequality is true at time-step $\tau$, cuboid $n$ and face $l$. When this happens $b^j_{\tau,n,l}$ becomes 1, otherwise $b^j_{\tau,n,l}=0$ and the constraint in Eqn. \eqref{eq:P1_x1} is valid with the use of a large positive constant $M$ as shown. Then the constraint in Eqn. \eqref{eq:P1_x2} uses the binary variable $\tilde{b}^j_{\tau,n}$ to count the number of times $b^j_{\tau,n,l}$ takes a value of one, and becomes active when $\sum_{l=1}^L b^j_{\tau,n,l}=6$ which signifies that agent $j$ resides inside the $n_\text{th}$ cuboid at time-step $\tau$. Finally, the constraint in Eqn. \eqref{eq:P1_x3} with the use of the binary variable $\hat{b}^j_{n}$ makes sure that the agent has no incentive in visiting the same cuboid multiple times during the current planning horizon.

Ideally, in this centralized multi-agent formulation we would like to have the following properties: a) a cuboid $\mathcal{C}_n$ that has been visited by some agent $i$ in the past, is not visited again by another agent $j$ in the future and, b) if agent $i$ plans to visit cuboid $\mathcal{C}_n$ in the future, then agent $j \ne i$ refrains from including cuboid $\mathcal{C}_n$ in its future plans, thus avoiding duplication of work. Properties a) and b) are accomplished by the constraint in Eqn. \eqref{eq:P1_6}. 
More specifically, each agent $j$ stores all visited cuboids in its local database $V^j_t \in \{0,1\}$, referred to as search-map hereafter, and uses this search-map in order to avoid visiting cuboids that it has visited in the past, thus avoiding the duplication of work. Therefore, $V^j_t(n)=1$ only when the cuboid $\mathcal{C}_n$ has been visited by the agent at time prior to $t$. In all other cases $V^j_t(n)=0$.
\noindent  The binary variable $\hat{b}^j_n$ indicates whether cuboid $n$ has been planned to be visited by agent $j$ during the next planning horizon i.e., $\hat{b}^j_{n} = 1, ~\text{iff} ~Hx^j_{t+\tau+1|t} \in \mathcal{C}_n, ~\tau \in [0,\ldots,T-1]$. Therefore, the inequality $y^j_{n} \le \hat{b}^j_{n} + V_t^j(n)$ provides no incentive for the agent to visit cuboids that have been visited in the past. In Eqn. \eqref{eq:P1_6}, the future plans of all other agents $i \ne j \in \mathcal{M}$ are denoted as $P_t^i(n)$ and defined as: $P_t^i(n)=1 \implies \exists ~ \tau \in [0,\ldots,T-1] : Hx^i_{t+\tau+1|t} \in \mathcal{C}_n$, otherwise $P_t^i(n)=0$.
%

\noindent As shown in (P1) by Eqn. \eqref{eq:P1_6} the past and future plans of all other agents $i \ne j \in \mathcal{M}$ denoted by $V_t^i(n)$ and $P_t^i(n)$ respectively are taken into account when deriving agent's $j$ plan by maximizing the binary variable $y^j_n$. There are four possible ways for activating $y^j_n$: a) the cuboid $\mathcal{C}_n$ has been planned to be visited by agent $j$ during the next planning horizon, which is indicated by $\hat{b}^j_n$, b) the cuboid has already been searched by agent $j$ as indicated by $V_t^j(n)$, c) the cuboid $\mathcal{C}_n$ is included in the future plans of some agent $i \ne j$ indicated by $P_t^i(n)$ and finally, (d) another agent $i \ne j$ has already searched cuboid $\mathcal{C}_n$ in the past as indicated by $V_t^i(n)$. Duplication of work occurs when more than one of $\{\hat{b}^j_n, V_t^j(n), V_t^i(n), P_t^i(n)\}, \forall i \ne j \in \mathcal{M}$ becomes active for a specific cuboid $\mathcal{C}_n$. However, even though the constraint in Eqn. \eqref{eq:P1_6} remains valid when more than one activations occurs, the less-than or equal sign effectively discourages such scenarios since the value of $y^j_n$ cannot exceed the maximum value of 1, and all the involved variables are binary. The centralized multi-agent formulation presented in problem (P1) with the availability of all the necessary information, optimally solves the joint multi-agent search planning problem, while avoiding the duplication of work by jointly considering the past and future plans of all agents.


\subsection{Distributed Control} \label{ssec:distributed}

A closer look at the centralized problem (i.e., P1), reveals the existence of coupled constraints as shown in Eqn. \eqref{eq:P1_6}. Consequently, the centralized formulation  requires at each time-step information (i.e., search maps and future plans) from all agents in order to produce the joint search plans by optimizing the objective shown in Eqn. \eqref{eq:centralized_mission_objective}. This is possible in the centralized version of the problem since all the information is available at the time of planning and moreover the problem is solved jointly among all agents on a central system. This ensures that the agents cooperate to minimize duplication of work. Although, the centralized version of the problem achieves optimality, it has several drawbacks: a) the computational complexity increases with the number of agents, b) it relies on the availability of information from all agents at every time-step and finally c) it does not accounts for failures on the central station where the planning process takes place.

The aforementioned drawbacks of the centralized system are alleviated in this work with the design of a distributed system \cite{Christofides2013}, however this comes at the cost of optimality. More specifically, in the proposed distributed control approach we drop the coupled constraints of Eqn. \eqref{eq:P1_6}, and the behavior of the centralized system is approximated as follows: At each time-step $t$, agent $j$ will compute a local search plan without considering the intentions of other agents, unless $\mathcal{N}_t^j \ne \emptyset$ i.e., agent $j$ receives the search plans of other agents $i \in \mathcal{N}_t^j$ inside its limited communication range. Subsequently, cooperative search plans are  generated in the scenario where two or more agents cooperate via communication, and exchange their future intentions. For this reason the constraints of Eqn. \eqref{eq:P1_6} are only approximated in the proposed distributed system as it is explained next in more detail. Nevertheless, the proposed distributed system offers an appealing tradeoff between optimality and computational complexity, as shown later in the evaluation.

Finally, the proposed distributed search planning framework is based on the following required key properties: First, the agents operate autonomously and in parallel with each other without the need for deliberative coordination.  The term coordination in this work refers to the ability of each agent to decide its own control inputs independently from other agents, and without relying on any specific execution order amongst the cooperative agents (e.g., sequential decision making/control procedures \cite{Tisdale2009,Richards2007}). In this work we would like to make sure that the mission will not be interrupted and it will be completed in the event where one or more agents need to exit the mission space. Constant communication between the agents should not be a requirement, rather the agents can opportunistically communicate and exchange information only when they are within communication range. Finally, the agents should cooperate and work towards improving the system-wide (i.e., collective) objective (i.e., searching all the objects of interest) while at the same time trying to minimize the duplication of work. 

\color{black}

\begin{figure}
	\centering
	\includegraphics[scale=0.29]{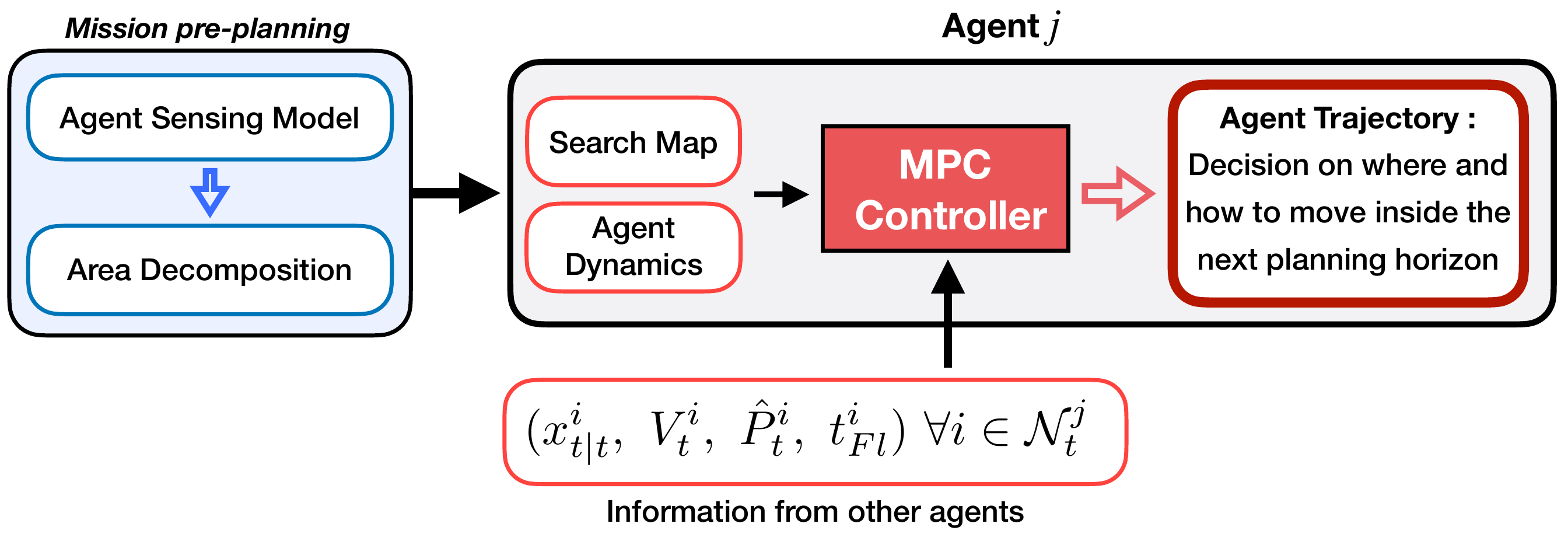}
	\caption{Overview of the proposed search planning framework.}	
	\label{fig:overview}
	\vspace{-5mm}
\end{figure}

In the distributed formulation of the problem we consider a team $\mathcal{M}$ of agents where each agent $j \in \mathcal{M}$ evolves inside a bounded surveillance area $\mathcal{A}$, according to the dynamics in Eqn. \eqref{eq:agent_dynamics}. Each agent $j$ exhibits a nominal flight time $\mathcal{T}^j$ and with probability $p^j_b(t)$ the agent's battery is depleted at time $t \le \mathcal{T}^j$. When a battery depletion event occurs i.e., $\mathcal{B}^j = 1$ the agent must exit the mission and return to its base station $\mathcal{G}^j \in \mathbb{R}^3$ for recharging as discussed in Sec. \ref{ssec:battery}. The subset of active agents (not recharging agents) at time $t$ is thus denoted as $\tilde{\mathcal{M}}_t \subseteq \mathcal{M}$.
Each agent exhibits a communication range $C_R$ for communicating and exchanging information with nearby agents. More specifically, agent $j$ receives the following information from all the neighboring agents $i \in \mathcal{N}_t^j$ at time $t$: a) agent's $i$ current state $x^i_{t|t}$, b) agent's $i$ search-map $V^i_t(n)$, c) agent's $i$ flight time $t^i_{Fl}$ and finally d) its future plan $\hat{P}^i_t(n)$. An overview of the proposed cooperative search planning framework is illustrated in Fig. \ref{fig:overview}.
We should point out here, that the future plan $\hat{P}^i_t(n)$ of some agent $i$ which is received by agent $j$ at time $t$ is not the most recent plan of agent $i$, since at the time of the communication agent $i$ has not yet generated its future plan for time $t$. Because the agents are synchronized, operate in parallel, and without coordination, at the time of communication the agents are not receiving the latest plans of their peers i.e., $\hat{P}^i_t(n) = P^i_{t-1}(n)$. We are going to refer to $\hat{P}^i_t(n)$ as the hypothetical plan of agent $i$ at time $t$ from the point of view of agent $j$. 
Furthermore, $\hat{P}^i_t(n) \in \{0,1\}, \forall n$ and is defined as $\hat{P}^i_t(n)=1, \text{if}~ \exists \tau \in [0,..,T-1] : Hx^i_{t+\tau|t-1} \in \mathcal{C}_n$. Hereafter, we will use the notation $\tau\hat{P}^i_t(n)$ to refer to the relative time $\tau$ in the planning horizon for which cuboid $n$ is planned to be visited i.e., $\hat{P}^i_t(n)=1$ by agent $i$. We can now describe the proposed distributed search planning formulation.


\subsubsection{Objective function}
The objective function of the centralized problem in Eqn. \eqref{eq:centralized_mission_objective} can be decomposed into several local objectives per agent as: $\sum_{j=1}^{|\tilde{\mathcal{M}}_t|} \mathcal{J}^j_\text{local}(X^j_t, U^j_t, Y^j)$,  
\noindent so that each active agent $j$ can independently optimize its local objective function $\mathcal{J}^j_\text{local}(X^j_t, U^j_t, Y^j)$ while at the same time the collective effort of the agents optimizes the system wide objective similarly to the centralized problem as discussed in Sec. \ref{ssec:centralized}.
The objective function of each agent becomes $\mathcal{J}^j_\text{local}(X^j_t, U^j_t, Y^j)$:
\begin{equation} \label{eq:distributed_mission_objective}
	\mathcal{B}_t^j \mathcal{J}^j_\text{recharge}(X^j_t) + (1-\mathcal{B}_t^j) \mathcal{J}^j_\text{search}(X^j_t, U^j_t, Y^j) 
\end{equation}

\noindent where $\mathcal{B}_t^j \in \{0,1\}$ indicates a battery depletion event which occurs with probability $p^j_b(t)$ and at which point the agent must return to its base station $\mathcal{G}^j \in \mathbb{R}^3$ for recharging by minimizing $\mathcal{J}^j_\text{recharge}(X^j_t) = \|Hx^j_{t+\tau+1|t}-\mathcal{G}^j\|^2_2$. On the other hand when $\mathcal{B}_t^j=0$, the agent $j$ optimizes its search planning objective $\mathcal{J}^j_\text{search}(X^j_t, U^j_t, Y^j)$ which is given by:
\begin{align} \label{eq:mission_search_objective}
    &\mathcal{J}^j_\text{search}(X^j_t, U^j_t, Y^j) =w_1 \|Hx^j_{t+\tau^\star+1|t}-x_j^{\star}\|^2_2 ~~+\\ \notag &
    ~~~~~~~~w_2\sum_{\tau=1}^{T-1} \|u^j_{t+\tau|t}-u^j_{t+\tau-1|t}\|^2_2 ~-~ w_3\sum_{n=1}^{N} r^j(n) y^j_{n}  &
\end{align}
\noindent where the first term guides the agent towards the unvisited cuboids, the second term minimizes the deviations between consecutive control inputs and finally the third term aims to maximize the number of cuboids that will be visited in the future as explained in Sec. \ref{ssec:centralized}. Specifically, $x^\star_j=c(\mathcal{C}_{n^\star_j})$ determines the centroid of the nearest cuboid with respect to agent $j$, with $n^\star_j$ given by:
\begin{equation} \label{eq:next_target}
n^\star_j = 
  \begin{cases} 
  \underset{n \in \tilde{N}^j}{\arg\min} \|Hx^j_{t+\tau^\star+1|t} - c(\mathcal{C}_n)\|_2, & \text{if}~ \mathcal{N}_t^j = \emptyset \\
   \underset{A^j}{\arg\min} \underset{i\in \{j \cup \mathcal{N}_t^j\}}{\sum}~ \underset{n \in \tilde{N}_t^j}{\sum} \Omega^j_{i,n} A^j_{i,n}, & \text{o.w.}   
  \end{cases}
\end{equation}
\noindent In particular, when agent $j$ is not in communication range with other agents i.e., $\mathcal{N}_t^j=\emptyset$, then agent $j$ moves greedily towards its nearest cuboid as shown above. However, when $\mathcal{N}_t^j \ne \emptyset$, agent $j$ receives the location of all other agents i.e., $Hx^i_{t|t}, i \in \mathcal{N}_t^j$, and hypothesizes what their next target (i.e., cuboid to be visited) will be. In other words agent $j$ adjusts its next target according to the hypothesized actions of the agents in its neighborhood. To do so, agent $j$ solves a local assignment problem where the objective is to find the cuboids that are likely to be visited next by the agents in the set $\mathcal{N}_t^j \cup j$. For this reason the cost matrix $\Omega^j_{i,n}$ is constructed locally at agent $j$, and populated with the distances between agent's $i$ location $Hx^i_{t|t}$ and every unvisited cuboid $n \in \tilde{N}_t^j$ ($\tilde{N}_t^j$ denotes the unvisited cuboids in agent's $j$ search-map). Then the objective is to find an assignment matrix $A^j$, which assigns the agents to the unvisited cuboids, where $A^j_{i,n} \in \{0,1\}$, and the sum of each row and column of $A^j$ does not exceeds the value of one. Once a solution is found agent $j$ keeps its assigned cuboid (i.e., extracts $n_j^\star$ from $A^j$) and discards all other results. 


 
 Finally, in the last term (i.e., $\sum_{n} r^j(n) y^j_{n}$),  $y^j_n$ is a binary decision variable which is activated whenever one of the following is true: a) cuboid $n$ has been visited by agent $j$ in the past, b) cuboid $n$ has been planned to be visited by agent $j$ in the current planning horizon or c) some agent $i \ne j$ has visited cuboid $n$ in the past and this information has already been communicated to agent $j$. The term $r^j(n) \in \{0,1\}$ is a reward term which is used to include or exclude cuboid $n$ from the planning process as we will explain next. Essentially, the notation $r^j(n)y^j_{n}$ indicates here whether the decision variable $y^j_{n}$ will be included in the optimization. Since there is no coordination between the agents, and because the agents operate in parallel it is highly likely that one or more agents (especially nearby agents) generate plans for the same cuboids. In addition, each agent $j$ with probability $p^j_b(t)$ will exit the mission space due to a depleted battery event. As a consequence, cuboids that have been planned to be visited by agent $j$ will be left unvisited in such events. Thus, the agents need to account for the above scenarios in an effort to increase the overall search planning performance and reduce the duplication of work. Let us assume that agent $j$ has received at time $t$ the hypothetical future plans of all nearby agents $i \ne j \in \mathcal{N}_t^j$ denoted as $\hat{P}^i_t(n), \forall n \in N$ where $N$ is the total number of cuboids in the environment. Alongside $\hat{P}^i_t(n)$ the agent has also received $\tau\hat{P}^i_t(n)$, and flight time $t^i_{Fl}$ for each agent $i \in \mathcal{N}_t^j$.

 With this information, agent $j$ first computes the probability that a particular cuboid $\mathcal{C}_n$ will not be visited by any agent that has made plans for it, due to the occurrence of battery depletion events. More specifically let $\mathcal{W}_t^j \subseteq \mathcal{N}_t^j$ to denote the subset of agents which have included cuboid $n$ in their plans transmitted to agent $j$ i.e., $\hat{P}^l_t(n) = 1, \forall l \in \mathcal{W}_t^j$ and let $\tau\hat{P}^l_t(n)$ to denote the relative time $\tau$ in the planning horizon for which agent $l \in \mathcal{W}_t^j$ is planning to visit cuboid $\mathcal{C}_n$. Agent $j$ computes the probability that agent $l \in \mathcal{W}_t^j$ will experience a battery depletion event before reaching cuboid $\mathcal{C}_n$ as:
 \begin{equation}
     p^l_F(n) = p^l_b(t^l_{Fl}+\tau\hat{P}^l_t(n)-1)
 \end{equation}
\noindent where $t^l_{Fl}+\tau\hat{P}^l_t(n)-1$ is the hypothesized arrival time of agent $l$ at cuboid $\mathcal{C}_n$. Subsequently, the probability of the event for which all agents $l \in \mathcal{W}_t^j$ fail to reach cuboid  $\mathcal{C}_n$ due to depleted batteries and agent $j$ does not runs out of battery during its planning horizon, is computed as:
\begin{equation} \label{eq:prob_fail}
    \hat{p}^j_F(n) =  \left(1-p^j_b(t+T)\right) \prod_{l=1}^{|\mathcal{W}_t^j|} p^l_F(n)
\end{equation}

\noindent The probability in Eqn. \eqref{eq:prob_fail} is computed by agent $j$ with information received from its communication neighborhood and allows the agent to determine whether a particular cuboid needs to be included in its future plans, given the hypothesized battery depletion events of other agents. The value of $\hat{p}^j_F(n) \in [0,1]$ indicates the probability with which agent $j$ should include cuboid $n$ in its future plans. 

As we have already mentioned, the plans received by agent $j$ from other agents inside its communication neighborhood are not necessarily up-to-date and could have been changed. For this reason agent $j$ takes into account the plans of other agents only with certain probability. More specifically, the expected number of agents $l \in \mathcal{W}_t^j$ that will reach cuboid $n$ during the next planning horizon can be computed as: $\hat{m}(n) = \sum_{l=1}^{|\mathcal{W}^j|} (1 - p^l_F(n))$. Based on the expected number of agents $\hat{m}(n)$ that agent $j$ hypothesizes that will reach cuboid $n$ in the future, agent $j$ includes cuboid $n$ in its future plans with probability which is given by:

\begin{equation} \label{eq:contigency}
\hat{p}^j_C(\hat{m}(n)) = 
  \begin{cases} 
   1 - \frac{1}{1+\alpha^j_2 e^{-\beta_2^j(\hat{m}(n)-\alpha_2^j)}}, & \text{if}~ p^j_b(t+T)>0.5  \\
   0, & \text{o.w}
  \end{cases}
\end{equation}

\noindent where $\beta_2^j$ and $\alpha_2^j$ are design parameters. In essence, Eqn. \eqref{eq:contigency} expresses the probability that a particular cuboid $n$ will be included in agent's $j$ plan conditioned on the expected number of agents that also plan to visit the same cuboid. This probability decreases with the expected number of agents that plan to visit a particular cuboid i.e., agent $j$ probabilistically refrains from visiting cuboid $n$ when a large number of agents are expected to visit $n$ as well. The reward $r^j(n)$ shown in Eqn. \eqref{eq:mission_search_objective} can now be defined as:
\begin{equation} \label{eq:reward}
r^j(n) = 
  \begin{cases} 
   1,  & \text{with probability}~ \max \{\hat{p}^j_F(n), \hat{p}^j_C(\hat{m}(n))\}  \\
   0, & \text{otherwise}
  \end{cases}
\end{equation}
\noindent Eqn. \eqref{eq:reward} is activated when $\forall~l \in \mathcal{W}_t^j : \hat{P}^l_t(n) = 1$. On the other hand, when no agent $l \in \mathcal{W}_t^j$ has included cuboid $n$ in its plans then $r^j(n) = 1$. Additionally, when $\mathcal{N}_t^j = \emptyset$ then $r^j(n) = 1, \forall n$.

To summarize each agent $j \in \tilde{\mathcal{M}}_t$ solves the distributed MPC problem shown in (P2) by optimizing their local objective function i.e., Eqn. \eqref{eq:distributed_mission_objective} with respect to their own control inputs $U^j_t$ and binary variables $Y^j$. In problem (P2) we assume that there are $q \in \mathcal{Q}$ objects of interest that need to be searched by the team of agents, and that each object of interest is searched when all $\mathcal{C}^q_n, n \in [1,..,N_q]$ cuboids are visited by at least one agent. Moreover, the agents must avoid collisions with all $\psi \in \Psi$ obstacles in the environment, including the objects of interest.

\begin{algorithm}
\begin{subequations}
\begin{align}
&\hspace*{-3.0mm}\textbf{Problem (P2)}: \texttt{Distributed MPC} &  \nonumber\\
& \hspace*{-3.0mm}~~~\underset{U^j_t,Y^j}{\min} ~\mathcal{J}^j_\text{local}(X^j_t, U^j_t, Y^j) &\label{eq:objective_P2} \\
&\hspace*{-3.0mm}\textbf{subject to} ~ \tau \in [0,\ldots,T-1] \textbf{:}  &\nonumber\\
&\hspace*{-3.0mm} x^j_{t+\tau+1|t} = \Phi x^j_{t+\tau|t} + \Gamma u^j_{t+\tau|t} - \Gamma u_g & \hspace*{1mm} \forall \tau \label{eq:P2_1}\\
&\hspace*{-3.0mm} x^j_{t|t} = x^j_{t|t-1} & \hspace*{1mm}\label{eq:P2_2}\\
&\hspace*{-3.0mm} x^j_{t+\tau+1|t} \in \mathcal{X} &\hspace*{1mm} \forall \tau \label{eq:P2_3}\\
&\hspace*{-3.0mm} |u^j_{t+\tau+1|t}| \le u_\text{max} &\hspace*{1mm} \forall \tau \label{eq:P2_4}\\
&\hspace*{-3.0mm} A_{q,n,l} H x^j_{t+\tau+1|t} ~~+ & \hspace*{1mm} \notag \label{eq:P2_5}\\
&\hspace*{-3.0mm} ~~~~~~~~~~(M-B_{q,n,l}) b^j_{\tau,q,n,l} \le M & \hspace*{1mm}\forall \tau,q,n,l \\
&\hspace*{-3.0mm} L \tilde{b}^j_{\tau,q,n} - \sum_{l=1}^L b^j_{\tau,q,n,l} \le 0 & \hspace*{1mm}\forall \tau, q, n \label{eq:P2_6}\\
&\hspace*{-3.0mm}  \hat{b}^j_{q,n} \le \sum_\tau \tilde{b}^j_{\tau,q,n} & \hspace*{1mm}\forall q, n \label{eq:P2_7}\\
&\hspace*{-3.0mm}  V_t^j(q,n) = V_t^j(q,n) + \sum_{i \ne j \in \mathcal{N}_t^j} V_t^i(q,n) & \hspace*{1mm}\forall q, n \label{eq:P2_8}\\
&\hspace*{-3.0mm} y^j_{q,n} \le \hat{b}^j_{q,n} + V_t^j(q,n) &\hspace*{1mm} \forall q, n \label{eq:P2_9}\\
&\hspace*{-3.0mm} A_{\psi,l} H x^j_{t+\tau+1|t} > B_{\psi,l} - M z^j_{\tau,\psi,l} &\hspace*{1mm} \forall \tau,\psi, l \label{eq:P2_10}\\
&\hspace*{-3.0mm} \sum_{l=1}^L z^j_{\tau,\psi,l} \le L-1 &\hspace*{1mm} \forall \tau,\psi \label{eq:P2_11}
\end{align}
\vspace{-0mm}
\end{subequations}
\end{algorithm}

\subsubsection{Constraints}

In problem (P2) each agent $j$ constructs its future trajectory $x^j_{t+\tau+1|t}$ over the rolling horizon $\tau \in [0,\ldots,T-1]$ of length $T$. The constraints in Eqn. \eqref{eq:P2_1} - \eqref{eq:P2_4} are due to the agent dynamical model. Then, the constraints in Eqn. \eqref{eq:P2_5} - \eqref{eq:P2_6} check whether the $n_\text{th}$ cuboid (i.e., $\mathcal{C}^q_n$), of the object of interest $q$, has been planned to be visited at time $t+\tau+1|t$ by the agent. To do this, we use the binary variables $b^j_{\tau,q,n,l}$ and $\tilde{b}^j_{\tau,q,n}$ where $l \in [1,..,L]$ denotes the cuboid faces. The constraint in Eqn. \eqref{eq:P2_7} discourages agent $j$ to include in its plans cuboid $n$ of the object of interest $q$ more than once during the planning horizon. Then the constraint in Eqn. \eqref{eq:P2_8} updates agent's $j$ search-map $V_t^j(q,n)$ with information received from other agents $i \in \mathcal{N}^j_t$. When agent $j$ has no agents inside its communication range $\mathcal{N}^j_t = \emptyset$ then $V_t^j(q,n)$ is not updated with information from other agents. Subsequently, the constraint in Eqn. \eqref{eq:P2_9} is used to give no incentive for agent $j$ to visit cuboid $n$ of the object of interest $q$, if $n$ has been visited in the past (by agent $j$ or any other agent which has exchanged information with agent $j$ at some point in time). The constraints in Eqn. \eqref{eq:P2_10} - \eqref{eq:P2_11} define collision avoidance constraints with the obstacles $\psi \in \Psi$.

\begin{figure*}
	\centering
	\includegraphics[scale=0.30]{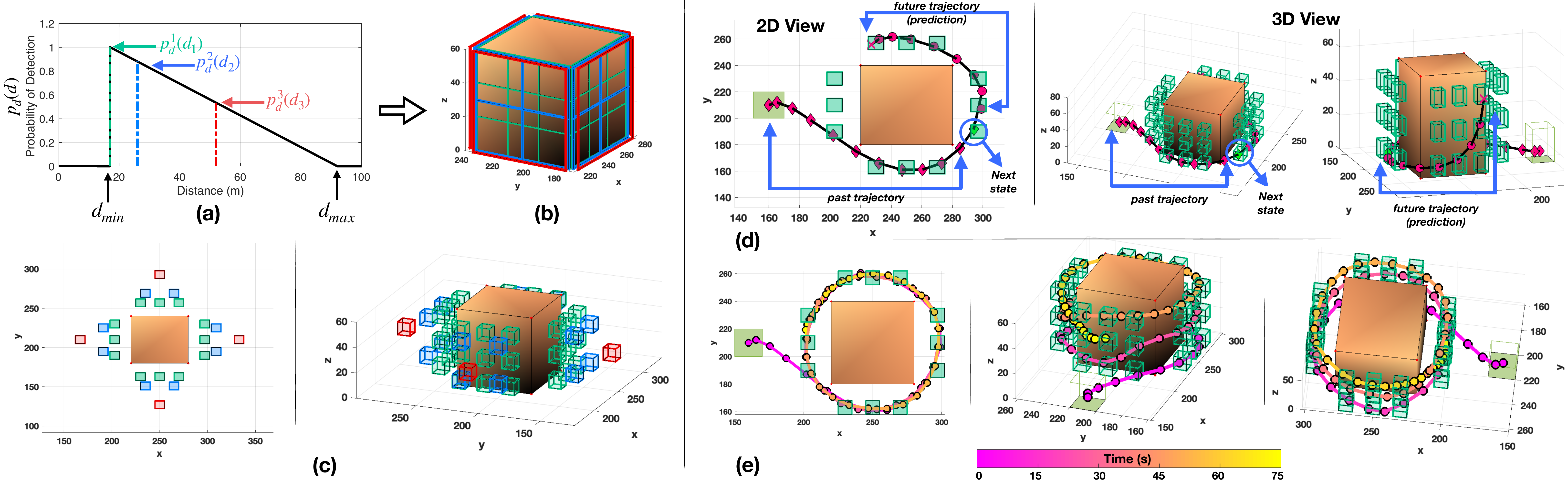}
	\caption{The figure illustrates: (a)-(c) the mission-preplanning step, (d)-(e) the generated 3D search plan for a single UAV agent.}	
	\label{fig:eval1}
	\vspace{-4mm}
\end{figure*}

\section{Evaluation} \label{sec:Evaluation}
The experimental setup used for the evaluation of the proposed system is as follows: The agent dynamics are expressed by Eqn. (\ref{eq:agent_dynamics}) with $\Delta T = 1$s, agent mass $m=3.35$kg and air resistance coefficient $\eta = 0.2$. The applied control input is bounded as $|u_t| \le 35 N$, the agent velocity is bounded within $|\dot{\text{x}}|\le 15$m/s, and the agent's position is bounded within the physical limits of the surveillance area $\mathcal{A}$. The agent FoV angle $\phi$ is set at 60deg. Simulations were conducted on an 3.5GHz dual core CPU running the Gurobi V9 MIQP solver. 

\subsection{Mission Pre-planning}
First we demonstrate the mission-preplanning step which is depicted in Fig. \ref{fig:eval1}(a)-(c). 
As we have discussed in Sec. \ref{sec:area_decomp}, in the mission pre-planning step, the mission control at the central station specifies the amount of search-effort required for efficiently searching an object of interest, which in this work is captured by the detection probability. In this scenario we assume that the profile of the detection probability is given by Eqn. \eqref{eq:pd} with $d_\text{min}=17$m and $d_\text{max}=90$m as shown in Fig. \ref{fig:eval1}(a). Subsequently, Fig. \ref{fig:eval1}(b) shows an example of the area decomposition step for 3 different detection probabilities i.e., $p_d^1(d_1), p_d^2(d_2)$ and $p_d^3(d_3)$, where $p_d^1(d_1)=1$ is the maximum detection probability, $p_d^2(d_2)=0.88$ and $p_d^3(d_3)=0.53$ indicated by the green, blue and red colors, respectively. Once the central station issues the required detection probability, the distance (i.e., $d_1, d_2$ or $d_3$) that the UAV agent must maintain with the object of interest is determined from Eqn. \eqref{eq:pd} and the FoV footprint is computed according to the agent's sensing model. In the illustrated scenario, $p_d^1(d_1), p_d^2(d_2)$ and $p_d^3(d_3)$ is achieved at distances $d_1=17\text{m}, d_2=26\text{m}$ and $d_3=52\text{m}$, respectively. The agent's FoV area for $d_1$ is approximately $20\text{m} \times 20\text{m}$, whereas for $d_2$ and $d_3$ the FoV sizes are  $30\text{m} \times 30\text{m}$ and $60\text{m} \times 60\text{m}$, respectively. 

Let us now assume that a large structure or building, with dimensions $60\text{m} \times 60\text{m} \times 60\text{m}$ as shown in Fig. \ref{fig:eval1}(b), is on fire and thus all its lateral faces must be searched to determine if there are trapped people inside.
Each one of the faces of the object of interest is decomposed into multiple cells according to the agent's FoV footprint, forming a grid as shown in the figure. For $p_d^1(d_1)$ each face  is decomposed into 9 cells shown in green color, for $p_d^2(d_2)$ each face is decomposed into 4 cells shown with blue color and finally for $p_d^3(d_3)$ the agent's FoV area captures the whole face of the object of interest as shown with red color in Fig. \ref{fig:eval1}(b) and thus one cell contains the entire face. Finally, depending on the required detection probability, for each cell, an artificial cuboid is generated and placed in front of the cell's center, at the distance which the agent's FoV area matches the area of the cell. This is depicted in 2D and 3D view in Fig. \ref{fig:eval1}(c), where the green, blue and red cuboids are associated with the detection probabilities $p_d^1(d_1),  p_d^2(d_2)$, and $p_d^3(d_3)$ respectively and are placed at distances $d_1=17\text{m}, d_2=26\text{m}$, and $d_3=52\text{m}$ respectively.

 To summarize, the UAV agent is required to pass from a total of 36 cuboids (i.e., 9 cuboids per face) in order to search the 4 faces (shown in Fig. \ref{fig:eval1}(b)) of the object of interest with a detection probability of $p_d^1(d_1)=1$. On the other hand, when the detection probability is set to $p_d^2(d_2)=0.88$ the agent needs to pass from 16 artificial cuboids and finally with a detection probability of $p_d^3(d_3)=0.53$ only 4 cuboids need to be visited. 

Once the mission-preplanning step is completed, the 3D search planning problem is transformed into an optimal control problem where the objective is to guide the UAV agents through all the generated artificial cuboids. Figure \ref{fig:eval1}(d)(e) shows the output of the rolling-horizon model predictive control formulation for a single agent, obtained from problem (P1) (problems (P1) and (P2) are equivalent in this case and produce the same result). The objective here is to search the object of interest discussed in the previous paragraph with the maximum detection probability i.e., $p_d^1(d_1)=1$. In this setting, the agent needs to visit a total of 36 cuboids around the object of interest shown in green color in Fig. \ref{fig:eval1}(d)(e). The agent's home depot is shown with a light green box and its initial state is $x_0 = [160, 200, 5, 0, 0, 0]$. The size of the surveillance region is $300\text{m} \times 300\text{m} \times 80\text{m}$, the planning horizon $T$ is set at 10 time-steps, the weights $w_1,w_2, \text{and}~w_3$ of the objective function in Eqn. \eqref{eq:centralized_mission_objective} are set to 0.0001, 0.0001, and 0.3 respectively and finally $\tau^\star = 3$. 

Figure \ref{fig:eval1}(d) shows the agent's trajectory at time-step 11. The agent's executed trajectory is denoted with red diamonds and the agent's predicted trajectory i.e., $x_{t+\tau+1|t}, t=11, \tau \in [0,..,9]$ is marked with red circles. As shown in the figure, the agent maximizes the number of cuboids to be visited within its planning horizon. Figure \ref{fig:eval1}(e) on the other hand, shows the final trajectory of the agent which took place over 75 time-steps. As it is shown the agent visits all the generated cuboids, forming a spiral trajectory. 
\begin{figure}
	\centering
	\includegraphics[width=\columnwidth]{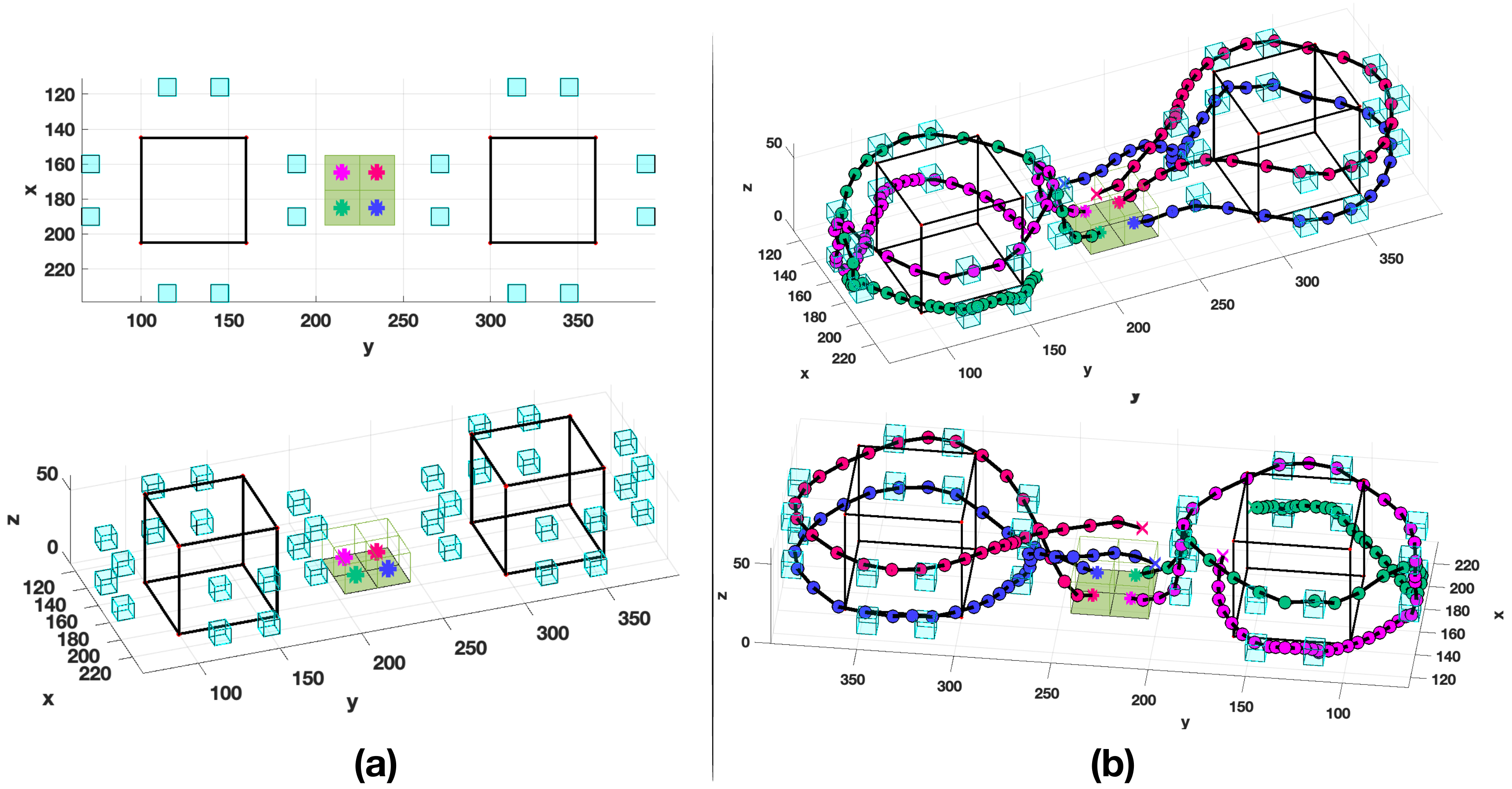}
	\caption{Distributed Search Planning with 4 cooperative UAV agents.}	
	\label{fig:eval2}
	\vspace{-5mm}
\end{figure}
\subsection{Distributed 3D Search Planning}\label{ssec:distributed_evaluation}
Next we analyze the performance of the proposed distributed 3D search planning approach. We begin our evaluation, with an illustrative scenario shown in Fig. \ref{fig:eval2}, where 4 agents are tasked to search 2 objects of interest with sizes $60\text{m} \times 60\text{m} \times 60\text{m}$ each. In this scenario the surveillance region has a size of  
$250\text{m} \times 400\text{m} \times 80\text{m}$ and the agents are required to search the objects of interest with a detection probability of $p_d^2(d_2)=0.88$, which results in the generation of 4 cuboids per face as shown in Fig. \ref{fig:eval2}(a). In order to search the 4 faces of each object of interest, as depicted in the figure, the agents need to visit 32 cuboids in total (colored in cyan). The 4 agents shown in purple, red, green and blue depart from their home depots as shown in the figure and execute the distributed MPC program shown in (P2) to produce the search trajectories illustrated in Fig. \ref{fig:eval2}(b). We should mention that for this experiment, the parameters $\alpha_1^j$ and $\beta_1^j$ of Eqn. \eqref{eq:battery_eq} have been set to 100 and 0.3 respectively for all agents. Similarly, the parameters $\alpha_2^j$ and, $\beta_2^j$ of Eqn. \eqref{eq:contigency} have been set to 2 and 0.5 respectively for all agents and the communication range $C_R$ was set to 100m. All the other parameters remain unchanged. The initial states of the agents are $x^1_0 = [166, 235, 5, 0, 0, 0]$, $x^2_0 = [185, 235, 5, 0, 0, 0]$, $x^3_0 = [165, 215, 5, 0, 0, 0]$, and $x^4_0 = [185, 215, 5, 0, 0, 0]$ and the planning horizon is $T=10$ time-steps. As it is shown in the figure, the agents work cooperatively to search the objects of interest in a distributed fashion. In particular we observe that the agents are divided into two teams i.e., the green-purple team and the blue-red team, with each team searching one object of interest. In this scenario all 32 cuboids are visited by the agents in 48 time-steps.

\begin{figure}
	\centering
	\includegraphics[scale=0.20]{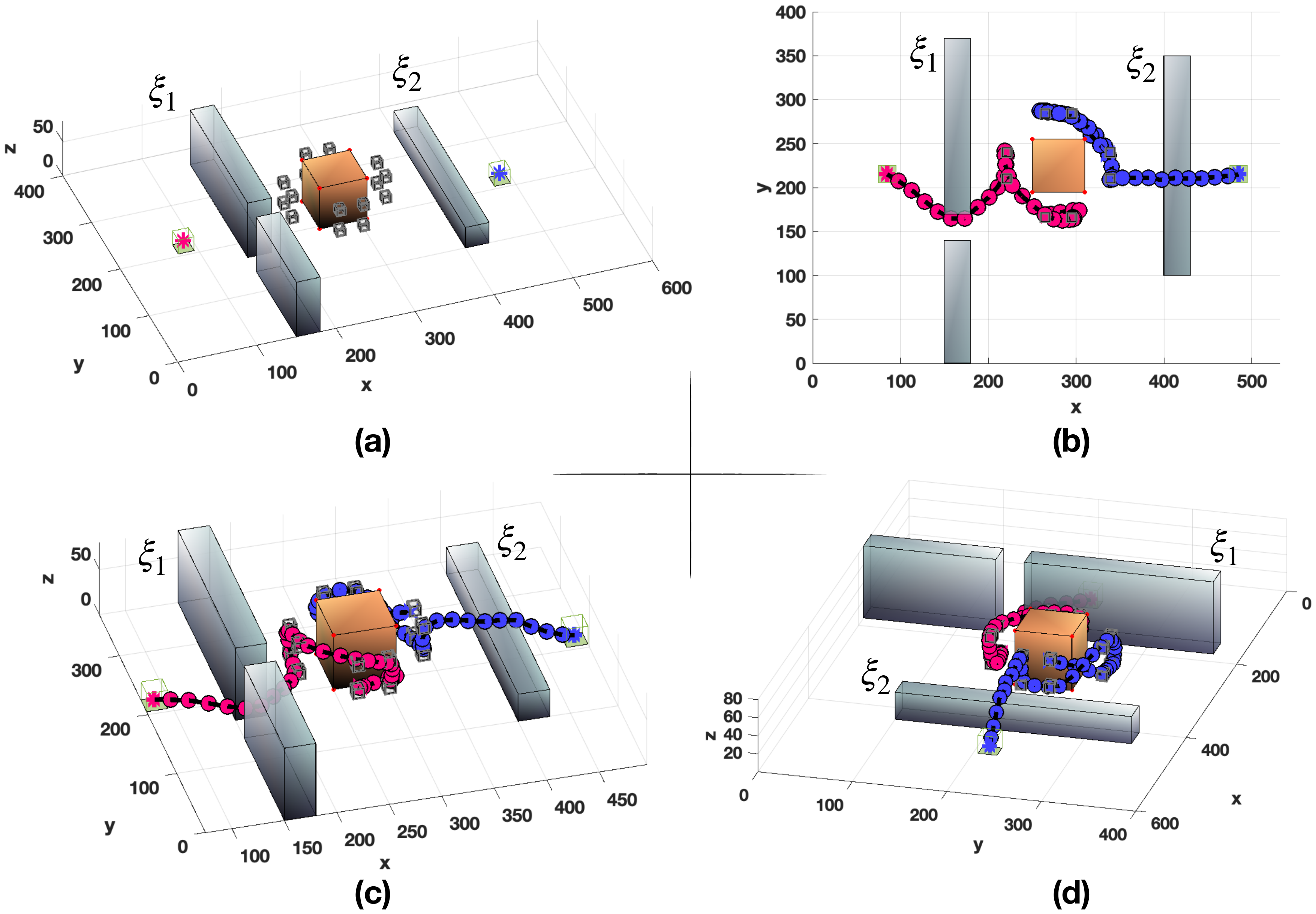}
	\caption{Searching with 2 cooperative UAV agents in the presence of obstacles.}
	\label{fig:eval3}
	\vspace{-5mm}
\end{figure}

The next experiment aims to demonstrate the cooperative behavior of the system in the presence of obstacles. This experiment is depicted in Fig. \ref{fig:eval3}, where 2 cooperative UAV agents, denoted with red and blue color, operate inside a surveillance region with dimensions $500\text{m} \times 400\text{m} \times 80\text{m}$. The agents initial state are $x^1_0 = [85, 215, 5, 0, 0, 0]$ and $x^2_0 = [485, 215, 5, 0, 0, 0]$ for the red and blue agents respectively. The agents collaborate in order to search a single object of interest (by visiting a total of 16 artificial cuboids) located between two obstacles $\psi_1$ and $\psi_2$ as as depicted in Fig. \ref{fig:eval3}(a). The height of obstacle $\psi_1$ is set at 80m (equal to the maximum height of the surveillance region), and the height of obstacle $\xi_2$ is 30m. As shown in the figure, each agent manages to search 2 of the object's faces, and thus in total 4 faces are searched (i.e., all 16 cuboids are visited by the agents as shown). More importantly, the agents avoid the obstacles in the environment in their effort to reach the object of interest. 

\begin{figure}
	\centering
	\includegraphics[width=\columnwidth]{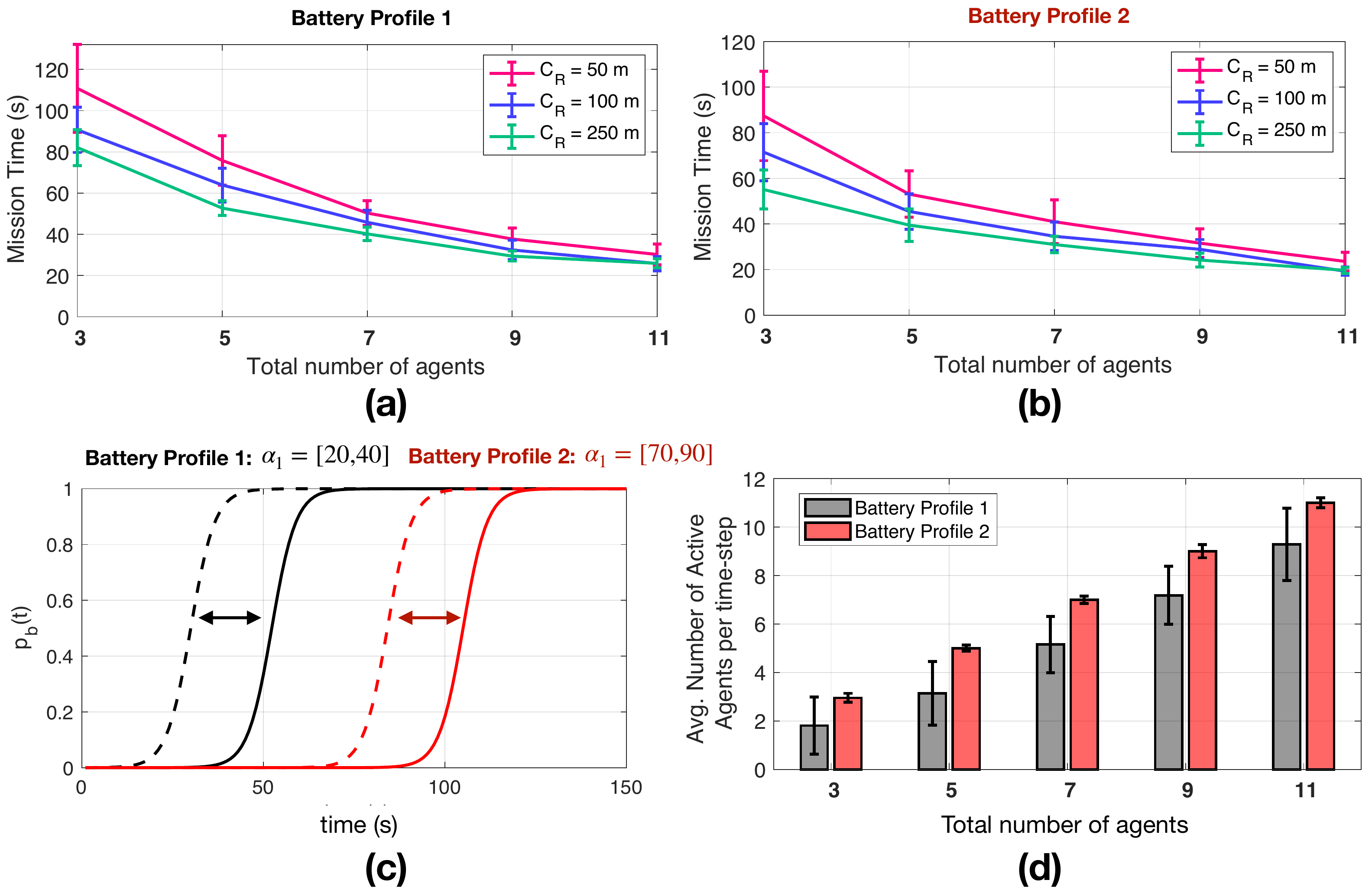}
	\caption{The figure shows the search planning performance for different parameter configurations of the proposed approach.}	
	\label{fig:eval4}
	\vspace{-5mm}
\end{figure}

The next series of experiments aims to investigate the impact of: a) the number of agents $|\mathcal{M}|$, b) the communication range $C_R$, and finally c) the parameters $\alpha_1$ and $\beta_1$ of the agent's battery profile i.e.,  Eqn. \eqref{eq:battery_eq}, on the mission completion time i.e., the amount of time required for searching all objects of interest. For this experiment we have used the environmental set-up shown in Fig. \ref{fig:eval2}, with two objects of interest of sizes $60\text{m} \times 60\text{m} \times 60\text{m}$ each, inside the surveillance region of size $400\text{m} \times 400\text{m} \times 80\text{m}$ and with a detection probability of $p_d=0.88$, which results in the generation of 32 artificial cuboids in total. In this test we have experiment with various parameter configurations as follows: the total number of available agents $|\mathcal{M}|$ varies in the set $\{3, 5, 7, 9, 11\}$, the communication range $C_R$ takes values in the set $\{50\text{m}, 100\text{m}, 250\text{m}\}$ and two different battery profiles settings have been used i.e., with the parameter $\alpha_1$ in the range $\alpha_1=[20,40]$ for battery profile 1 and $\alpha_1=[70,90]$ for battery profile 2. The parameter $\beta_1$ is kept fixed at $\beta_1=0.3$. The agent recharging time $t_R$ is sampled uniformly from the interval $[5,10]$. We have conducted 50 Monte Carlo (MC) trials for each parameter combination, where we randomly initialize the agents inside the surveillance region and we let the system (i.e., problem (P2)) to run, logging the mission completion time and the number of active agents per time-step. The averaged results for the different configurations are illustrated in Fig. \ref{fig:eval4}. More specifically, Fig. \ref{fig:eval4}(a) shows the average mission completion time for different agent team sizes and various communication ranges for battery profile 1. For this experiment, the battery profile parameter $\alpha_1$, for each agent is sampled uniformly within the interval $[20,40]$. On the other hand, in Fig. \ref{fig:eval4}(b) the same configuration scenario is simulated for battery profile 2, in which $\alpha_1$ is sampled uniformly within the interval $[70,90]$. The conditional probability distributions of the two battery profiles are shown in Fig. \ref{fig:eval4}(c) with black and red colors, for profile 1 and 2, respectively. As we can observe from Fig. \ref{fig:eval4}(a) and Fig. \ref{fig:eval4}(b), the average mission time decreases as the number of agents increases. Additionally, these results also show the impact of the communication range on the performance of the system. As the communication range increases, the cooperation between the agents also increases which results in improved mission execution times. Interestingly, we can observe that a large communication range primarily benefits small teams, for which the agents are sparse and scattered within the surveillance region. Fig. \ref{fig:eval4}(b), shows a similar behavior for battery profile 2. However, in this scenario the battery depletion events are not as frequent, compared to the battery depletion events obtained with battery profile 1. As a result, the average number of active agents per time-step, participating in the mission is larger, which improves the mission execution times as shown. Figure \ref{fig:eval4}(d), shows the average number of active agents per time-step for the two different battery profiles. As it is shown, the frequent battery depletion events caused with battery profile 1, makes the number of agents that participate in the mission to fluctuate significantly, which can potentially decrease the system's performance.  Nevertheless, the results show the flexibility of the proposed distributed search planning approach to cope with a dynamically varying number of agents. 

\begin{figure}
	\centering
	\includegraphics[width=\columnwidth]{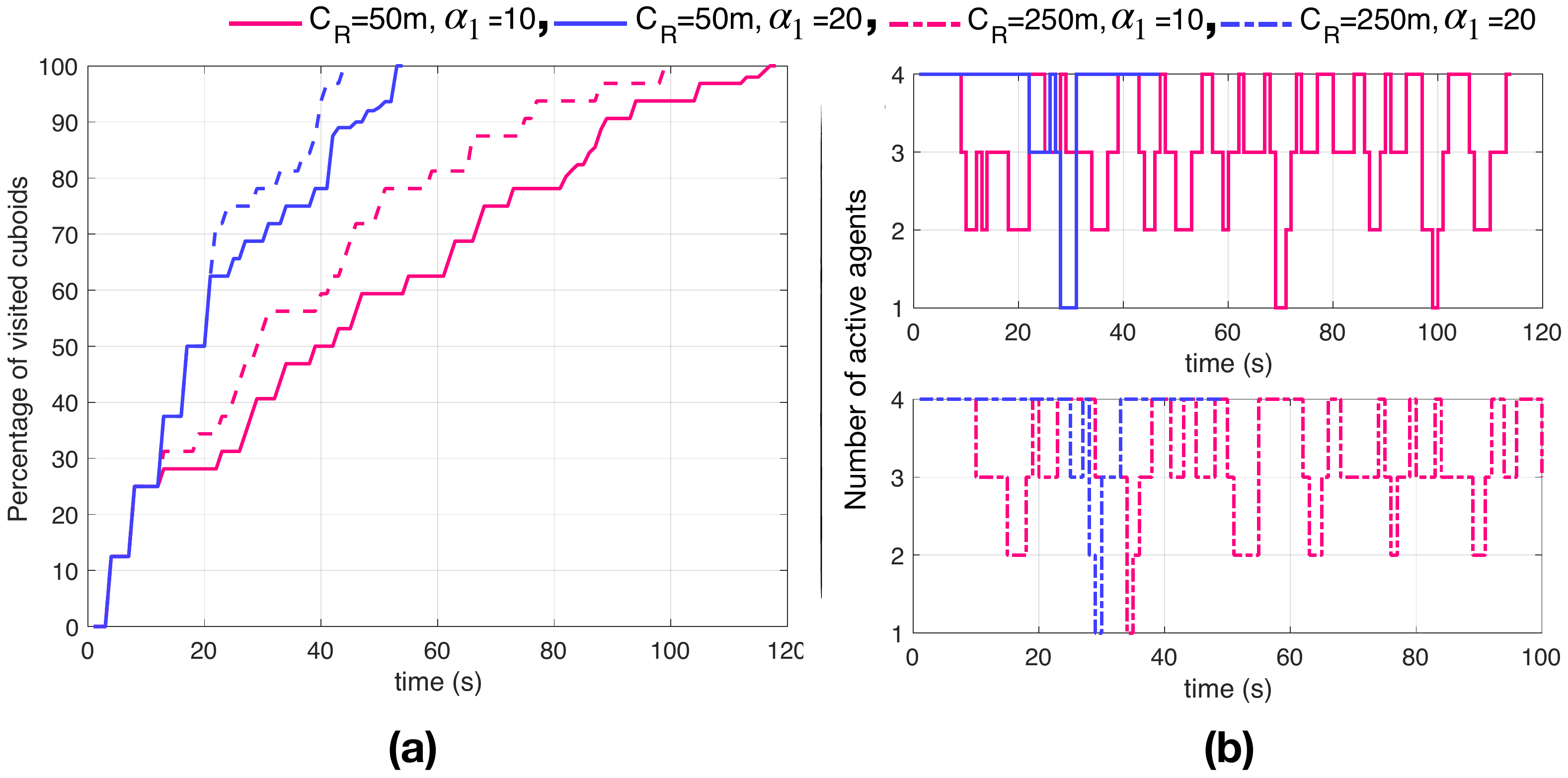}
	\caption{The figure shows the percentage of visited cuboids over time and the number of active agents in each time-step, for different parameter configurations, for a search planning scenario with 4 agents.}	
	\label{fig:eval5}
    \vspace{-5mm}
\end{figure}

The next experiment, aims to demonstrate more clearly the effect of the battery profile and the communication range, on the performance of the system. In this experiment, we used the setup shown in Fig. \ref{fig:eval2}, with 4 agents initialized at $x^1_0 = [166, 235, 5, 0, 0, 0]$, $x^2_0 = [185, 235, 5, 0, 0, 0]$, $x^3_0 = [165, 215, 5, 0, 0, 0]$, and $x^4_0 = [185, 215, 5, 0, 0, 0]$, and with the objects of interest as shown in the figure. The agent recharging time $t_R$ is sampled uniformly within the interval $[1,5]$ and the rest of the parameters remain unchanged. We run the system with the following configurations a) $(C_R=50\text{m}, \alpha_1=10)$, b) $(C_R=50\text{m}, \alpha_1=20)$, c) $(C_R=250\text{m}, \alpha_1=10)$ and, d) $(C_R=250\text{m}, \alpha_1=20)$, and we monitor the percentage of visited cuboids over time as shown in Fig. \ref{fig:eval5}(a). Figure \ref{fig:eval5}(b) shows the number of active agents in each time-step, for the four configurations. As it is shown in Fig. \ref{fig:eval5}(a) with the red and blue solid lines, the system's performance increases dramatically with the reduction of battery depletion events. This is also evident by the number of active agents shown in Fig. \ref{fig:eval5}(b). When $\alpha_1=10$, the agents enter and exit the mission space very frequently as shown by red solid and dotted lines, which causes delays in the mission execution time. Here it is also evident the importance of the communication range on the performance of the system. The increased communication range i.e., dotted lines, significantly improves the mission execution times, as shown in the figure.

\begin{figure}
	\centering
	\includegraphics[scale=0.3]{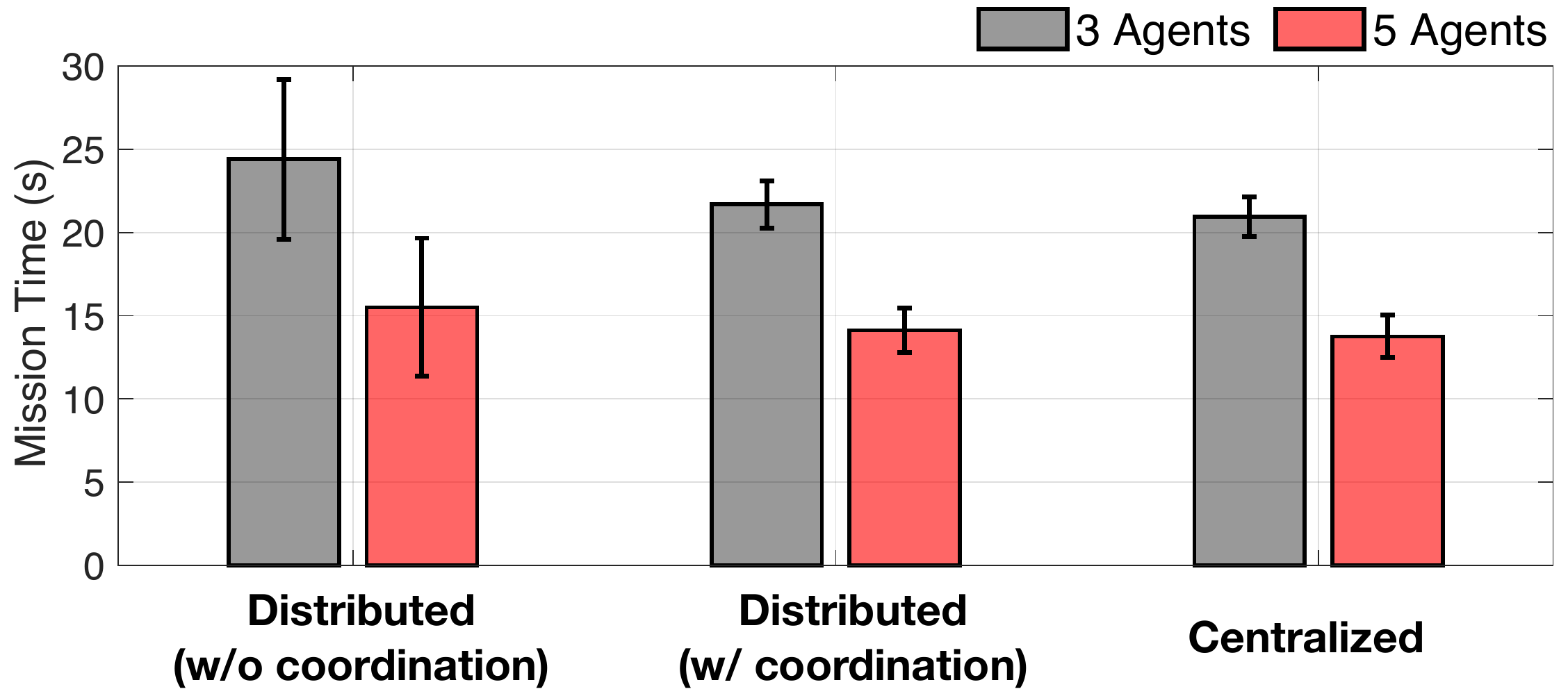}
	\caption{Performance comparison of the proposed distributed search planning approach with competing techniques.}	
	\label{fig:eval6}
    \vspace{-5mm}
\end{figure}

The next experiment compares the performance of the proposed distributed search planning approach, with the centralized formulation of the problem discussed in Sec. \ref{ssec:centralized}, and with the distributed planning framework presented in \cite{Richards2007} which requires coordination between the agents. Specifically, in \cite{Richards2007} the agents execute their plans in a sequential fashion one after the other.  In order to evaluate the 3 approaches discussed above we have used the following simulation setup: We have generated a surveillance region of size $300\text{m} \times 300\text{m} \times 80\text{m}$, with one object of interest of size $60\text{m} \times 60\text{m} \times 60\text{m}$, centered at $(x,y) = (175,130)$, with 16 artificial cuboids. The agents have a communication range of 430m, and no battery depletion events occur during searching. We conduct the experiment with $\alpha_1^j = 100$, $\beta_1^j = 0.3$, $\alpha_2^j=1.5$, $\beta_2^j=1$, and all other parameters as set as previously.
We have conducted 50 MC trials, where 3 and 5 agents are uniformly spawned inside the surveillance region. Fig. \ref{fig:eval6} shows the average mission completion time (i.e. the time required so that all cuboids are searched) for the 3 approaches. For the case of 3 agents, Fig. \ref{fig:eval6}, shows an average mission completion time of 24.4 seconds for the proposed distributed approach without coordination, approximately 21.8 seconds for the distributed framework with coordination, and 20.9 seconds for the centralized approach. Similar results have been obtained for the case of 5 agents, as shown in Fig. \ref{fig:eval6}. In summary, the centralized approach outperforms both distributed approaches in terms of mission completion time, and the coordination between the agents seems to provide a slight advantage over the proposed approach. However, both competing techniques require constant communication amongst the agents in order to produce plans. The centralized approach requires at each time-step all information to be transmitted to a central station. Similarly, the distributed approach with coordination is not flexible in terms of communication since it requires information exchange among all pairs of agents at each time-step, which also prohibits this technique from operating at high frequencies for large number of agents. In addition, the centralized approach does not scales well with the number of agents as shown in the next section, and the distributed approach with coordination is not robust to agent failures as opposed to the proposed approach. Consequently, the performance of the proposed distributed search planning framework seems reasonable for a system which does not require any form of coordination between the agents. Additionally, in the proposed framework the agents can enter and exit the mission space at random times and can communicate opportunistically, properties which fit with the application scenario studied in this work. 


\color{black}
\subsection{Computational Complexity}
The main factors that drive its computational complexity of the problem (P1) are a) the length of the planning horizon $T$, b) the number of agents $|\mathcal{M}|$ which are involved in searching and c) the number of cuboids $N$ that need to be searched. This is also evident by the number of binary variables required by the mixed integer quadratic program (MIQP) as shown in Eqn. \eqref{eq:P1_7}.  On the other hand, the computational complexity of the distributed search planning approach shown in Problem (P2), only grows with the length of the planning horizon $T$ and, with the number of cuboids $N$ that need to be searched. 


In this test we have run the centralized and distributed formulation of the proposed approach for $3$ and $7$ agents, with planning horizon lengths of $3$ and $7$ time-steps. For each (Number of Agents, Horizon Length) combination $20$ trials were conducted, with the agents being randomly initialized inside a square area of size 20m by 20m, located between two objects of interest as shown in Fig. \ref{fig:eval2}. In this experiment the number of cuboids $N$ that need to be searched was kept constant with $N=32$. The rest of the parameters were set according to the first paragraph of Sec. \ref{ssec:distributed_evaluation}.
Table~I summarizes the results of this experiment in terms of the execution time (i.e., the time required by the solver to find the optimal solution). In particular, Table~I shows the average execution time for each combination of the parameters for the centralized and distributed controllers. The results verify our previous discussion and show that the computational complexity of the centralized formulation does not scale well with the number of agents, as compared to the proposed distributed approach. 

\renewcommand{\arraystretch}{1.0}
\begin{table}\small
\label{tbl:tbl1}
\caption{Computational Complexity}
\begin{center}
  \begin{tabular}{|c|c|c|c|}
   \hline
    \multicolumn{4}{|c|}{Avg. Execution Time (sec)} \\
    \hline
    \# Agents & Horizon Length & Centralized & Distributed\\
    \hline\hline
    3 & 3  & 0.8969  & 0.0022 \\ \hline
    5 & 3  & 2.1220  & 0.0025 \\ \hline   
    3 & 7 & 11.5956   & 0.0756 \\ \hline
    5 & 7 & 56.3984  & 0.0787 \\ \hline
  \end{tabular}
\end{center} 
\vspace{-7mm}
\end{table}
\color{black}

\section{Conclusion} \label{sec:conclusion}
We have proposed, a novel distributed search-planning framework, where a dynamically varying number of autonomous agents cooperate in order to search multiple objects of interest in 3D. The proposed distributed model predictive control (MPC) approach allows the generation of cooperative search trajectories over a finite planning horizon and enables the agents to operate without coordination, optimizing their plans on-line for maximizing the collective search planning performance. 

\section*{Acknowledgments}
This work is funded by the Cyprus Research and Innovation Foundation under Grant Agreement EXCELENCE/0421/0586 (GLIMPSE), by the European Union’s Horizon 2020 research and innovation programme under grant agreement No. 739551 (KIOS CoE), and from the Government of the Republic of Cyprus through the Cyprus Deputy Ministry of Research, Innovation and Digital Policy.

\bibliographystyle{IEEEtran}
\bibliography{IEEEabrv,main}

\vspace*{0 mm}
\begin{IEEEbiography}[{\includegraphics[width=1in,height=1.25in,clip,keepaspectratio]{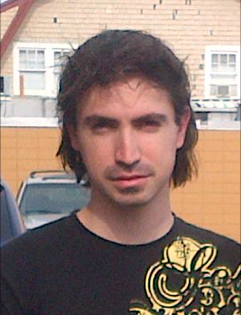}}]%
{Savvas Papaioannou} received the B.S. degree in Electronic and Computer Engineering from the Technical University of Crete, Greece in 2011, the M.S. degree in Electrical Engineering from Yale University, USA, in 2013, and the Ph.D. degree in Computer Science from the University of Oxford, UK in 2017. He is currently a Research Associate with the KIOS Research and Innovation Center of Excellence, University of Cyprus. His research interests include multi-agent and autonomous systems, state estimation and control, multi-target tracking, probabilistic inference, and intelligent unmanned-aircraft vehicle (UAV) systems and applications. Dr. Papaioannou is Reviewer for various conferences and journals within the IEEE and ACM associations.
\end{IEEEbiography}

\vspace*{0 mm}
\begin{IEEEbiography}[{\includegraphics[width=1in,height=1.25in,clip,keepaspectratio]{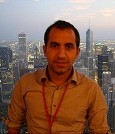}}]%
{Panayiotis Kolios} is a Research Associate at the KIOS Research and Innovation Center of Excellence at the University of Cyprus. He received his B.Eng and Ph.D degrees in Telecommunications Engineering from King's College London in 2008 and 2011, respectively. His interests focus on both basic and applied research on networked intelligent systems. Some examples of such systems include intelligent transportation systems, autonomous unmanned aerial systems, and the plethora of cyber-physical systems that arise within the Internet of Things. Particular emphasis is given to emergency response aspects in which faults and attacks could cause disruptions that need to be effectively handled. He is an active member of IEEE, contributing to a number of technical and professional activities within the Association. 
\end{IEEEbiography}

\vspace*{0 mm}
\begin{IEEEbiography}[{\includegraphics[width=1in,height=1.25in,clip,keepaspectratio]{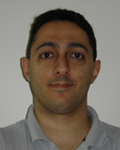}}]%
{Theocharis Theocharides} is an Associate Professor in the Department of Electrical and Computer Engineering, at the University of Cyprus and a faculty member of the KIOS Research and Innovation Center of Excellence where he serves as the Research Director. Theocharis received his Ph.D. in Computer Engineering from Penn State University, working in the areas of low-power computer architectures and reliable system design, where he was honored with the Robert M. Owens Memorial Scholarship, in May 2005. He has been with the Electrical and Computer Engineering department at the University of Cyprus since 2006, where he directs the Embedded and Application-Specific Systems-on-Chip Laboratory. His research focuses on the design, development, implementation, and deployment of low-power and reliable on-chip application-specific architectures, low-power VLSI design, real-time embedded systems design and exploration of energy-reliability trade-offs for Systems on Chip and Embedded Systems. His focus lies on acceleration of computer vision and artificial intelligence algorithms in hardware, geared towards edge computing, and in utilizing reconfigurable hardware towards self-aware, evolvable edge computing systems. His research has been funded by several National and European agencies and the industry, and he is currently involved in over ten funded ongoing research projects. He serves on several organizing and technical program committees of various conferences (currently serving as the Application Track Chair for the DATE Conference), is a Senior Member of the IEEE and a member of the ACM. He  is currently an Associate Editor for the ACM Transactions on Emerging Technologies in Computer Systems, IEEE Consumer Electronics magazine, IET's Computers and Digital Techniques, the ETRI journal and Springer Nature Computer Science. He also serves on the Editorial Board of IEEE Design \& Test magazine.  
\end{IEEEbiography}

\newpage

\vspace*{0 mm}
\begin{IEEEbiography}[{\includegraphics[width=1in,height=1.25in,clip,keepaspectratio]{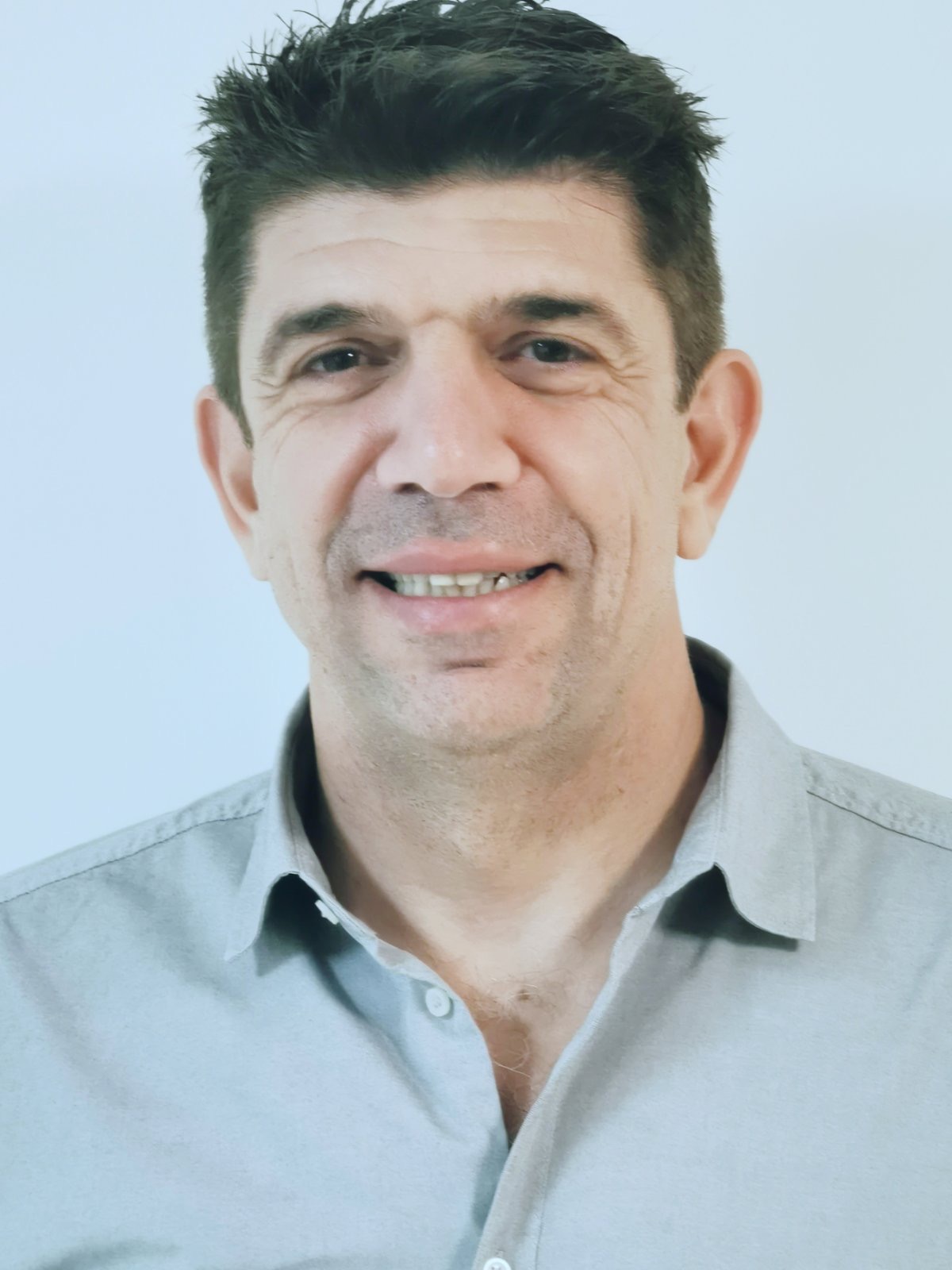}}]%
{Christos G. Panayiotou} is a Professor with the Electrical and Computer Engineering (ECE) Department at the University of Cyprus (UCY). He is also the Deputy Director of the KIOS Research and Innovation Center of Excellence for which he is also a founding member.  Christos has received a B.Sc. and a Ph.D. degree in Electrical and Computer Engineering from the University of Massachusetts at Amherst, in 1994 and 1999 respectively. He also received an MBA from the Isenberg School of Management, at the aforementioned university in 1999. Before joining the University of Cyprus in 2002, he was a Research Associate at the Center for Information and System Engineering (CISE) and the Manufacturing Engineering Department at Boston University (1999 - 2002). His research interests include modeling, control, optimization and performance evaluation of discrete event and hybrid systems, intelligent transportation systems, cyber-physical systems, event detection and localization, fault diagnosis, wireless, ad hoc and sensor networks, smart camera networks, resource allocation, and intelligent buildings. His research has been funded with more than 45 million euros from national and European funding agencies, governmental organizations and private companies.

 Christos has published more than 275 papers in international refereed journals and conferences and is the recipient of the 2014 Best Paper Award for the journal Building and Environment (Elsevier).  He is an Associate Editor for the IEEE Transactions of Intelligent Transportation Systems, the Conference Editorial Board of the IEEE Control Systems Society, the Journal of Discrete Event Dynamical Systems and the European Journal of Control. During 2016-2020 he was Associate Editor of the IEEE Transactions on Control Systems Technology. He held several positions in organizing committees and technical program committees of numerous international conferences, including General Chair of the 23rd European Working Group on Transportation (EWGT2020), and General Co-Chair of the 2018 European Control Conference (ECC2018). He has also served as Chair of various subcommittees of the Education Committee of the IEEE Computational Intelligence Society.
\end{IEEEbiography}

\vspace*{0 mm}
\begin{IEEEbiography}[{\includegraphics[width=1in,height=1.25in,clip,keepaspectratio]{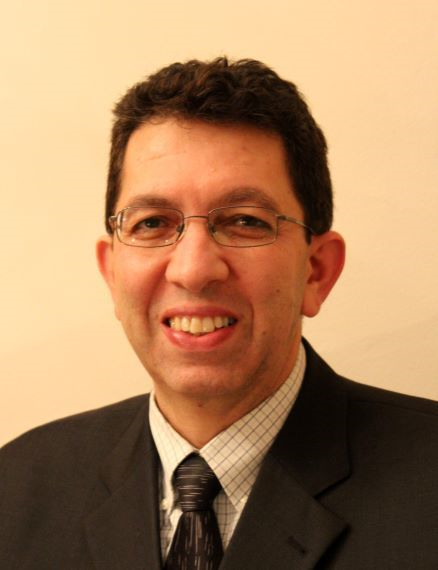}}]%
{Marios M. Polycarpou} is a Professor of Electrical and Computer Engineering and the Director of the KIOS Research and Innovation Center of Excellence at the University of Cyprus. He is also a Member of the Cyprus Academy of Sciences, Letters, and Arts, and an Honorary Professor of Imperial College London. He received the B.A degree in Computer Science and the B.Sc. in Electrical Engineering, both from Rice University, USA in 1987, and the M.S. and Ph.D. degrees in Electrical Engineering from the University of Southern California, in 1989 and 1992 respectively. His teaching and research interests are in intelligent systems and networks, adaptive and learning control systems, fault diagnosis, machine learning, and critical infrastructure systems. Dr. Polycarpou has published more than 350 articles in refereed journals, edited books and refereed conference proceedings, and co-authored 7 books. He is also the holder of 6 patents.
Prof. Polycarpou received the 2016 IEEE Neural Networks Pioneer Award. He is a Fellow of IEEE and IFAC and the recipient of the 2014 Best Paper Award for the journal Building and Environment (Elsevier). He served as the President of the IEEE Computational Intelligence Society (2012-2013), as the President of the European Control Association (2017-2019), and as the Editor-in-Chief of the IEEE Transactions on Neural Networks and Learning Systems (2004-2010). Prof. Polycarpou serves on the Editorial Boards of the Proceedings of the IEEE, the Annual Reviews in Control, and the Foundations and Trends in Systems and Control. His research work has been funded by several agencies and industry in Europe and the United States, including the prestigious European Research Council (ERC) Advanced Grant, the ERC Synergy Grant and the EU Teaming project.
\end{IEEEbiography}

\flushbottom
\balance

\end{document}